\documentclass[twocolumn]{aastex631}

\newcommand\kms{\rm{km\ s}^{-1}}

\shorttitle{Dust continuum and [\ion{C}{2}] $158\mu\rm{m}$ at 140 pc resolution in a $z=6.6$ quasar }
\shortauthors{Meyer et al.}

\graphicspath{{./}{figures/}}

\begin{document}

\title{Pushing ALMA to the limit: 140 pc resolution observations of a z=6.6 quasar-galaxy merger resolve strikingly different morphologies of dust continuum and [\ion{C}{2}] 158$\mu\rm{m}$ emission}

\author[0000-0001-5492-4522]{Romain A. Meyer}
\affiliation{Department of Astronomy, University of Geneva, Chemin Pegasi 51, 1290 Versoix, Switzerland}
\affiliation{Max Planck Institut f\"ur Astronomie, K\"onigstuhl 17, D-69117, Heidelberg, Germany}
\author[0000-0001-9024-8322]{Bram Venemans}
\affiliation{Leiden Observatory, Leiden University, PO Box 9513, 2300 RA Leiden, The Netherlands}
\author[0000-0002-9838-8191]{Marcel Neeleman}
\affiliation{National Radio Astronomy Observatory, 520 Edgemont Road, Charlottesville, VA, 22903, USA}
\affiliation{Max Planck Institut f\"ur Astronomie, K\"onigstuhl 17, D-69117, Heidelberg, Germany}
\author[0000-0002-2662-8803]{Roberto Decarli}
\affiliation{INAF – Osservatorio di Astrofisica e Scienza dello Spazio di Bologna, via Gobetti 93/3, I-40129, Bologna, Italy}
\author[0000-0003-4793-7880]{Fabian Walter}
\affiliation{Max Planck Institut f\"ur Astronomie, K\"onigstuhl 17, D-69117, Heidelberg, Germany}


\begin{abstract}
We present $0\farcs026$ $(140\ \rm{pc})$ resolution ALMA observations of [\ion{C}{2}] $158\ \mu\rm{m}$ and dust continuum emission of the $z=6.6$ quasar J0305--3150, resolved over $\sim 300-400$ independent resolution elements. The dust continuum emission is compact with $\sim 80\%$ recovered within $r<0\farcs3$ $(1.6\ \rm{kpc})$, whereas the [\ion{C}{2}] emission profile is composed of a central Gaussian ($r<0\farcs4$, i.e. $<2.2\ \rm{kpc}$) and an extended component (detected up to $\sim 10\ \rm{kpc}$ at 
$>3\sigma$). We infer a direct contribution of the quasar to the observed 260\ \rm{GHz} continuum $S_{\nu,\rm{QSO}} / S_{\nu,\rm{QSO+Host}} \lesssim 1\%$. We report the detection of FIR-detected star-forming clumps with $r<200 \ \rm{pc}$ and properties similar to that of rest-frame UV-optical clumps reported in the literature.
The $200\ \rm{pc}$ resolved [\ion{C}{2}]/FIR ratio follows the global relation with the FIR surface brightness established in low- and high-redshift galaxies, even at the quasar location.
We find that dust continuum is emitted in regions of $\sim0\farcs02-0\farcs04$ consistent with the size of photo-dissociation regions (PDR), whereas $50\%$ of the [\ion{C}{2}] originates from larger physical scales ($\theta \gtrsim 2"$). The large-scale [\ion{C}{2}] emission presents a velocity gradient aligned with a nearby companion with perturbed kinematics, and misaligned with the kinematics of the small-scale emission. The absence of significant [\ion{C}{2}] emission by structures with physical scale $\lesssim 1\ \rm{kpc}$ implies that [\ion{C}{2}] emission is not produced in dense PDR located at the boundary of Giant Molecular Clouds. We argue instead that [\ion{C}{2}] is produced in low-density PDRs in the interstellar medium and diffuse \ion{H}{1} gas tidally-stripped during the ongoing merger.
\end{abstract}


\section{Introduction}

The existence of luminous quasars powered by supermassive black holes (SMBH) with $M_{\rm{BH}} \gtrsim 10^8\, \rm{M}_\odot$  at $z>6$ poses a number of  challenges to models of early SMBH formation and growth \citep[e.g.,][]{DeRosa2014,Banados2018,Mazzucchelli2017,Wang2021}. To match the black hole masses observed in high-redhift quasars only $\sim 1\ \rm{Gyr}$ after the Big Bang, most simulations and theoretical models invoke sustained (super-)Eddington accretion rates \citep[see, e.g.,][for comprehensive reviews]{Haiman2004,Overzier2009,Volonteri2010,Latif2016,Inayoshi2020,Volonteri2021}. As this rapid growth can only be fuelled by gas from the host galaxy, studying the properties and environment of $z>6$ quasar host galaxies is crucial to understand the formation of the first SMBHs.

Observations in the (sub)-millimeter regime with ALMA and NOEMA/PdBI have provided an unhindered view of the interstellar medium of $z>6$ quasar host galaxies. The extreme star formation rates (SFR) of quasar host galaxies \citep[$\rm{SFR}\simeq 100-1000\ M_\odot\ \rm{yr}^{-1}$, e.g.][]{Maiolino2005,Walter2009,Cicone2015, Decarli2018,Shao2017, Shao2022, Venemans2016, Venemans2017, Venemans2018,Venemans2020, Wang2011, Wang2013, Wang2016b}, $1-2$ orders of magnitude higher than UV-selected galaxies at the same redshift \citep[e.g.][]{Inami2022,Schouws2022}, indicate that they are still accreting gas and building up their stellar mass. This is supported by observations showing that the most luminous $z>6$ quasars and AGNs host over-massive BHs compared to the local $M_* - \sigma$ relation  (e.g. \citealt{Neeleman2021,Yue2023, Harikane2023}, but see \citealt{Ding2023}). Quasar host galaxies have also been shown to often have large gas reservoirs likely fuelling their rapid growth \citep[$\sim 10^{10}\ M_\odot$, e.g.][]{Walter2003,Wang2011,Shao2017, Venemans2017, Feruglio2018, Novak2019, Meyer2022,Decarli2022,Kaasinen2024}. What remains unclear however is the details of the physical processes enabling rapid accretion of gas from the cosmic web or via mergers, as well as the gas pathways linking the circumgalactic medium to the central black hole. Detecting and resolving the large-scale gas reservoirs and the ISM of high-redshift quasar hosts is therefore necessary to understand the build-up of the first massive galaxies and the rapid growth of their central SMBHs.

One major limitation of current studies is the lack of observations at high resolution. Although existing $\sim 1\ \rm{kpc}$ observations can provide global properties of the galaxies \citep[e.g.][]{RWang2019, Venemans2020,Pensabene2020, Neeleman2021,Shao2022}, exploring the detailed morphology and physical conditions of $z>6$ quasar hosts requires higher spatial resolution \citep[e.g.][]{DiMascia2021,Lupi2022}. However, observations of $z\sim 6$ quasar host galaxies at resolutions $<500\ \rm{pc}$ remain rare, with only four observations available in the literature. Recently, \citet[][]{Walter2022,Meyer2023,Neeleman2023} have presented [\ion{C}{2}] observations of three of these quasars at resolution of $200-300\ \rm{pc}$. The three objects all show compact morphologies with most flux emitted within $r<1 \rm{kpc}$, but very different kinematics (a rotating disk, a dispersion-dominated galaxy and a warped spiral). By contrast, the similar-resolution ($\sim 0\farcs076$ / $\sim 400\ \rm{pc}$) imaging of the fourth quasar, J0305--3150, has been shown to have extended [\ion{C}{2}] and continuum emission up to $\sim 2-10\ \rm{kpc}$, complex kinematics with potential [\ion{C}{2}] cavities \citep[][]{Venemans2019}, and intense merger activity with no less than four close ($r=2.0-11.0\ \rm{kpc}$) companion galaxies. This makes J0305--3150 a unique laboratory to study how mergers can affect the growth of quasar hosts galaxies and help fuel the central SMBHs. Indeed, mergers are thought to be an efficient mechanism to channel gas into the center of the galaxy, thus triggering starbursts and enabling rapid central SMBH growth \citep[e.g.][]{Springel2005, DiMatteo2005, Hopkins2008, Hirschmann2010, Dubois2014, Kulier2015}.

 In this work, we revisit the $z=6.6$ quasar J0305--3150 with new observations of the [\ion{C}{2}] and continuum emission at an unprecedented resolution of $0\farcs026$ ($140\ \rm{pc}$ at $z=6.6$), pushing the capabilities of ALMA with its most extended configuration. We detail the observations and the data reduction in Section \ref{sec:obs_red}, and show that distribution of [\ion{C}{2}] and dust is strikingly different. In particular, $50\%$ of the [\ion{C}{2}] emission is over-resolved by our longest baseline observations, which is not the case for the continuum. In Section \ref{sec:c2_pdr} we study the morphology of the cold dust continuum and the [\ion{C}{2}]-emitting gas, showing that J0305--3150 harbours extended [\ion{C}{2}] emission, and that the [\ion{C}{2}] and dust emission are not emitted at the same physical scales. In Section \ref{sec:kinematics} we further show that the kinematics of the small-scales and large-scale [\ion{C}{2}] emission differ, and conclude that the latter traces tidally-stripped gas due to a recent merger. We discuss the resolved ISM properties of J0305--3150 (SFRD, [\ion{C}{2}] and depletion timescales) in Section \ref{sec:dust_SFRD} before  concluding in Section \ref{sec:conclusion}.

Throughout this work, we use a concordance cosmology with $H_0 = 70\ \rm{km\ s}^{-1}\ \rm{Mpc}^{-1}$, $\Omega_m=0.3$, $\Omega_\Lambda= 0.7$, implying a scale of 5.406 kpc arcsec$^{-1}$ at the redshift of J0305--3150. We account for residual-correction in the aperture-integrated flux densities \citep[e.g.][]{Jorsater1995,Walter1999,Novak2019} using the latest version of \textit{interferopy} \citep[v1.0.2][]{interferopy}. 

\section{Observations and data reduction}
\label{sec:obs_red}
\begin{figure*}
    \centering
    \includegraphics[width=\textwidth]{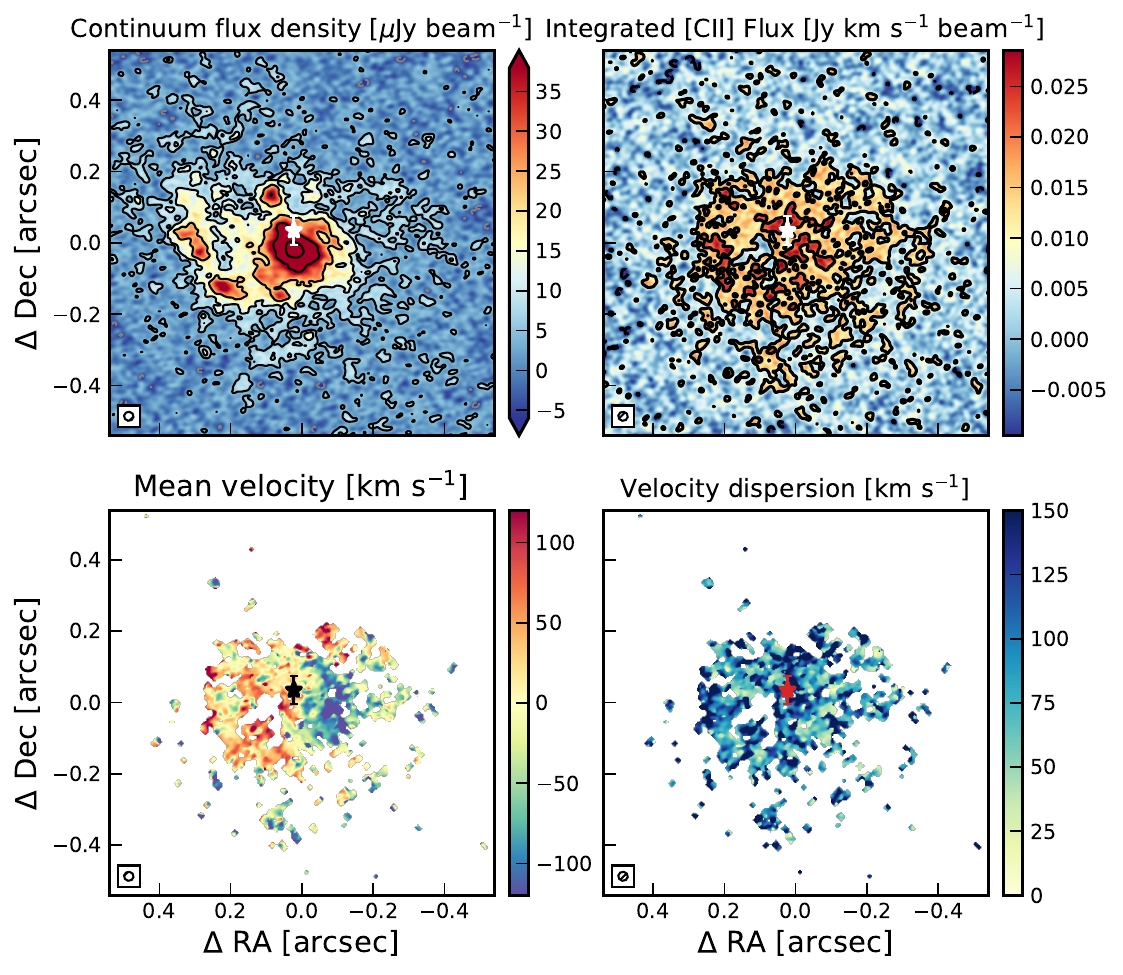}
    \caption{Top: FIR continuum at $\sim 260\ \rm{GHz}$ and velocity-integrated [\ion{C}{2}] emission of the high--redshift quasar J0305--3150, based on all available ALMA observations, including earlier ‘lower'--resolution data published by \citet{Venemans2016,Venemans2019}. The contours start at $\pm2\sigma$ and increase in powers of two. Positive contours are shown in full dark lines, and negative ones in dashed grey. The synthesized beam is plotted in the bottom left corner of each plot. The GAIA-corrected optical position of the quasar is shown with a star \citep[][]{Venemans2019}. Bottom: Mean velocity and velocity dispersion map of the [\ion{C}{2}] emission, computed using a Gaussian fit in pixels where [\ion{C}{2}] is detected at $>2\sigma$. The mask for the moment 1 and 2 maps is given by the moment 0 $2\sigma$ contours, with five rounds of binary erosion and binary dilation to remove small structures due to noise (see text for more details).} 
    \label{fig:fig1}
\end{figure*}

We observed J0305--3150 with ALMA in configuration C43-9/10 in 2021 August and September (Cycle 7) for a total on-source integration time of 15.5h. J0321-3122 was used as the phase calibrator and J0522-3627 as the flux and bandpass calibrator. The spectral windows were tuned to the observed frequency of [\ion{C}{2}] at $z=6.6$ and matched to the spectral setups of the previous ALMA observations \citep[Cycle 2, 3 and 5][]{Venemans2016,Venemans2019}. The combined ALMA dataset spans configurations C-5 to C-10, with baseline lengths ranging from $14\rm{m}$ up to $\sim16\ \rm{km}$, corresponding roughly to observed scales of $30"$ to $\sim15\ \rm{mas}$ (see further Table \ref{tab:observations_summary}). The new C43-9/10 data has a nominal resolution of $0\farcs018$ (e.g. $97\ \rm{pc}$ at $z=6.6$), but throughout the paper we use the combined dataset with a resolution of $0\farcs026$ (see further further Appendix \ref{app:baselines} for more details on the baseline distribution). The datasets were calibrated using the ALMA Pipeline for each respective cycle \citep{Hunter2023}. The calibrated datasets were combined, imaged and cleaned with CASA 6.2.1.7 \citep{THECASATEAM2022}. 

We produce continuum and [\ion{C}{2}] maps, as well as a full datacube of the [\ion{C}{2}] emission. The continuum map was produced using the three spectral windows which did not contain [\ion{C}{2}] emission. The [\ion{C}{2}] spectral line cube was constructed by subtracting the continuum in the uv plane using a first order polynomial fitted to the channels without line emission (defined as $|\nu -\nu_{\rm{[CII]}}|>2.5\times\ \rm{FWHM(\rm{[CII]})}$) in the side band covering the [\ion{C}{2}] line (e.g in the ranges $248.75<\nu [\rm{GHz}]<249.225$ and $249.90<\nu [\rm{GHz}]<249.62$ covered by the spectral window containing [\ion{C}{2}] and the directly adjacent one. We used the FWHM and line center from the low-resolution data \citep[][]{Venemans2019}, which are in agreement with that retrieved from the C43-9/10 data alone. We also generated a velocity-integrated [\ion{C}{2}] spectral line image from the continuum-subtracted data set by combining all channels within $<1.2\times \rm{FWHM}\ (\rm{e.g.}, <270\ \rm{km\ s}^{-1})$ of the center of the [\ion{C}{2}] line in the lower sideband \citep[][]{Venemans2019}. The full data cube, the [\ion{C}{2}] spectral line image and the dust continuum maps were imaged using natural weighting \textit{multiscale} cleaning down to $2\sigma$. We use a pixel size of $3\ \rm{mas}$ and a channel width of $30\ \rm{MHz}\ (36\ \rm{km\ s}^{-1}$) for the [\ion{C}{2}] datacube (the native resolution is $\sim4.4\ \rm{km}\ \rm{s}^{-1}$). We resolve the dust continuum and [\ion{C}{2}] in $\sim 400$ and $\sim 300$ independent resolution elements detected at $>2\sigma$. 

\begin{table}
    \centering
    \begin{tabular}{cccccc}
         Project ID  & $\theta_{res}$ & $\theta_{MRS}$ & Cont. rms & $t_{\rm{int}}$ \\
         & & & [mJy] & [s] \\
         2013.1.00273.S & $0\farcs208$ & $2\farcs210$ & 0.0409 & 907\\
         2015.1.00399.S & $0\farcs158$ & $2\farcs891$ & 0.0172 & 2298 \\
         2017.1.01532.S & $0\farcs029$ & $0\farcs834$ & 0.0147 & 13245 \\
         2019.1.00746.S & $0\farcs017$  & $0\farcs431$ & 0.0129 & 55865 \\
    \end{tabular}
    \caption{Summary of the ALMA data used for in this analysis. The angular resolution, maximum recoverable scale (MRS) and continuum sensitivity are estimates provided by the ALMA archive from the baselines (see further the ALMA technical handbook \url{https://almascience.nrao.edu/documents-and-tools/cycle10/alma-technical-handbook}). }
    \label{tab:observations_summary}
\end{table}

The synthesized beam is $0\farcs0267\times0\farcs0254$ for the [\ion{C}{2}] line map (the dust continuum beam size is $\sim 5\%$ smaller due to a better uv plane coverage). For the [\ion{C}{2}] datacube, the rms noise in the spectral line data cube is $\sigma = 36.1\  \rm{\mu Jy\ beam}^{-1}$ (in $30$\ MHz ($36\ \kms$) channels). The continuum map has an rms $\sigma =  3.33\ \mu \rm{Jy\ beam}^{-1}$, and the velocity-integrated [\ion{C}{2}] image $\sigma =  4.9\times 10^{-3}\ \rm{Jy\ beam}^{-1}\ \rm{km\ s}^{-1}$ (where we give the rms measured in the image, multiplied by $270\ \rm{km\ s}^{-1} / 0.84$ to provide the line sensitivity, \citep[see e.g.][for an explanation of the 0.84 factor]{Novak2019}). We checked that the synthesized beam is well behaved. Indeed, the positive (negative) sidelobe are $<3\%(-2\%)$ the maximum amplitude of the beam, and the beam pattern is symmetric and quasi circular (see further Appendix. \ref{app:baselines}). We conclude that none of the features analysed later in this paper are caused by the beam pattern.

The moment maps (mean velocity and dispersion) were computed by fitting a Gaussian line to each non-masked voxel using the \textit{Qubefit} library \citep[][]{qubefit}. It is important to note that moment images showing a $0\farcs9\times0\farcs9$ field of view, such as presented in this work, are produced from a 3D cube with $180\times180\times50=1.62\times10^6$ voxels. Assuming that the noise follows a Gaussian distribution, we would expect $\sim 4350$ voxels at SNR$>3$. In turn, this creates tens of SNR$>2$ (groups of) pixels in the velocity-integrated [\ion{C}{2}] map. Most will be smaller than the beam size (a few pixels only), and can be thus easily distinguished from true emission features. To do so, we add an extra step to the standard SNR$>2$ [\ion{C}{2}] emission mask of by using several rounds of binary erosion and dilation \footnote{Binary erosion and dilation are a standard morphological operations typically used in image processing. In a binary 0-1 pixel map, a pixel with value 1 is transformed to 0 if any neighbouring pixel is 0 following binary erosion. Conversely, binary dilation transforms any 0 pixel to 1 if any neighbouring pixel is 1. Successive N rounds of erosion and dilation thus close small holes with size N whilst preserving larger structures.}. To summarize, the mask is first defined as all pixels where the [\ion{C}{2}] emission is detected at $>2\sigma$ and is then modified by applying three rounds of binary erosion followed by three rounds of binary dilation to remove small structures due to noise. The number of erosion/dilation operations is chosen such that the number of pixels eroded/dilated corresponds roughly to the FWHM the beam.

We present the continuum and the velocity-integrated [\ion{C}{2}] maps as well as the mean velocity and dispersion field of the [\ion{C}{2}] line in Figure \ref{fig:fig1}. The dust continuum and [\ion{C}{2}] emission show a strikingly different morphology. Whereas the dust is concentrated within the central $r\sim 0\farcs1$ region, the [\ion{C}{2}] emission has a flatter profile over a much wider area ($r\sim 0\farcs3$). We leave a technical comparison of the recovered flux densities in dust and [\ion{C}{2}] between our observations including the C43-9/10 data and that presented in \citet[][]{Venemans2019}, respectively to the interested reader in Appendix \ref{app:flux_comparisons}. In summary, the dust emission is recovered within errors even when considering the C43-9/10 data, whereas $\approx 50\%$ of the [\ion{C}{2}] emission is lost compared to \citet[][]{Venemans2019}. Since the C43-9/10 antenna configuration becomes insensitive to emission on scales $r\gtrsim 0\farcs7$ / $3.7$ \rm{kpc} (see Appendix \ref{app:flux_comparisons}), this strongly suggests that, besides their different global morphology, the [\ion{C}{2}] and dust are emitted in regions with different sizes and physical properties. 

The velocity and dispersion maps of J0305-3150 (Figure \ref{fig:fig1}, bottom) present a velocity gradient consistent with a rotating disk as discussed in \citet[][]{Venemans2019, Venemans2020}. However, at such high resolution the velocity and dispersion profiles becomes highly irregular and potentially over-resolved, and we do not attempt to fit a thin disk model to improve on the dynamical mass constraint. \citet[][]{Venemans2019} already pointed out the irregular kinematics of J0305-3150 and suggested jets or cavities as potential explanations. Our data do not support such hypotheses, and we defer to Section \ref{fig:kinematics_2_scales} for an in-depth discussion of the kinematics in this object taking into account the peculiar [\ion{C}{2}] profile.

\section{The physical sizes of \texorpdfstring{[\ion{C}{2}]}{[CII]} and dust continuum emitting regions}
\label{sec:c2_pdr}
\subsection{The spatial extent of \texorpdfstring{[\ion{C}{2}]}{[CII]} and dust continuum emission}

\begin{figure}
    \centering
    \includegraphics[width=0.5\textwidth]{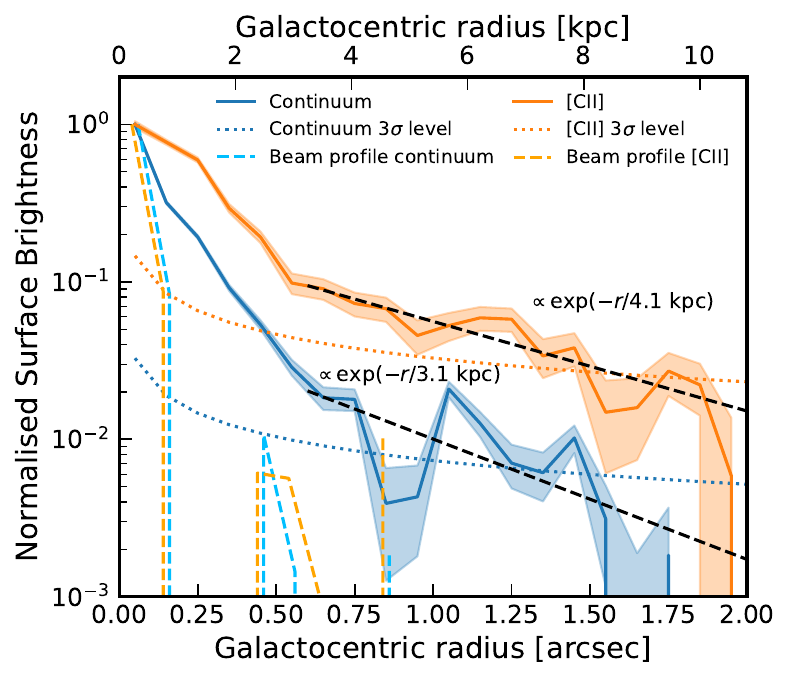}
    \caption{Normalised [\ion{C}{2}] and dust continuum surface brightness profiles, computed from all available Band 6 [\ion{C}{2}] data (see text). Profiles using only the latest $97\ \rm{pc}$--resolution (Cycle 7, C43-9/10) data are shown as dashed lines and lighter colors. Both profiles are much more extended than the synthesized beam profile of the observations (dashed lines - offset for clarity). We also show the best-fit exponentials to the emission profiles (at $r>0\farcs6$) . }
    \label{fig:cii_dust_profile}
\end{figure}

\begin{figure*}
    \centering
    \includegraphics[width=\textwidth]{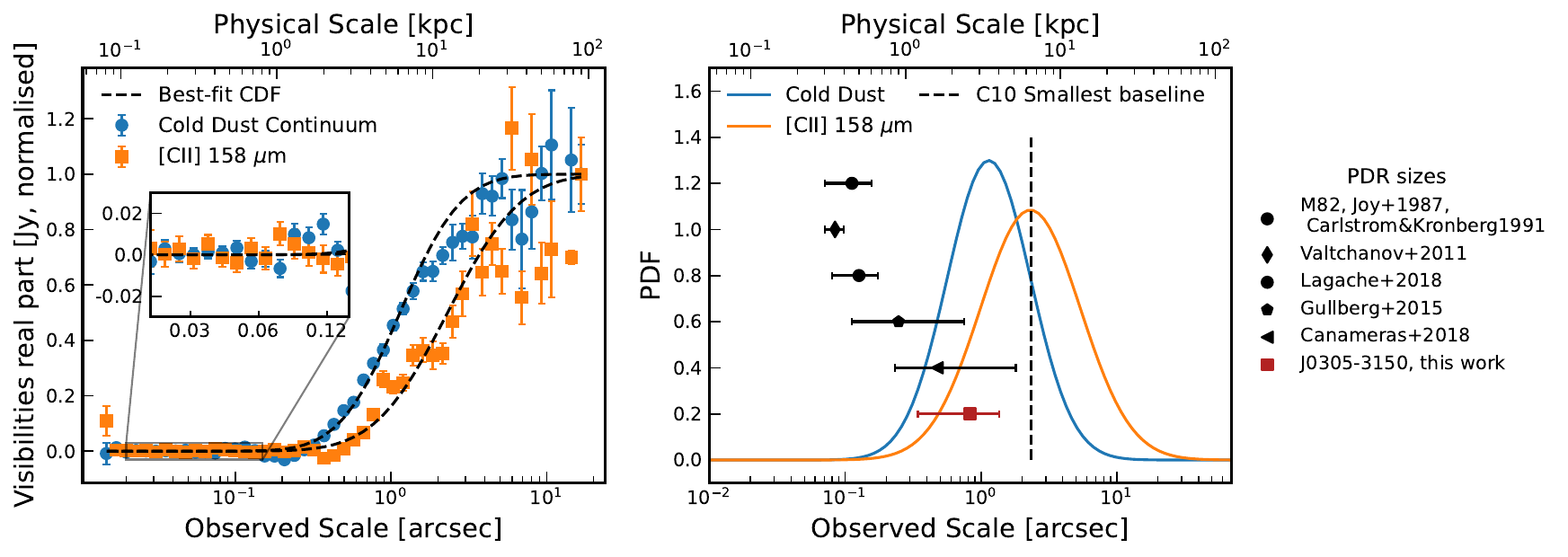}
    \caption{\textbf{Left panel:} Real part of the visibilities for the continuum and [\ion{C}{2}] emission, averaged as a function of baseline length converted to physical scales. The errorbars represent the standard deviation in each bin, and the visibilities are normalised to the first datapoint (\textit{uv} distances $(15\pm1)\rm{m}$). The best-fit cumulative distribution functions are shown in dashed black. \textbf{Right panel:} Best-fit probability density function for the continuum and [\ion{C}{2}] emission as a function of the logarithm of the physical scale observed. The maximum recoverable scale of the ALMA C10 observations is shown in dashed black. We also indicate the  photo-dissociation regions (PDR) sizes where [\ion{C}{2}] is thought to originate found in simulations and observations at $0\lesssim z\lesssim 5$ (black markers and errors), and provide a similar estimate for J0305--3150 in dark red.  \nocite{Carlstrom1991} }
    \label{fig:visiblities_scales}
\end{figure*}
The most striking feature of Fig.\ \ref{fig:fig1} is the difference in morphology between the dust continuum and [\ion{C}{2}] emission. Whilst both dust and [\ion{C}{2}] are clearly resolved and not point-source-like, the smooth [\ion{C}{2}] emission is clearly more extended than the compact and clumpy continuum emission. We first quantify the respective extension of [\ion{C}{2}] and continuum using the azimuthally-averaged surface brightness profiles shown in Figure \ref{fig:cii_dust_profile}. The surface brightness profiles are computed by summing the flux in annuli of $0\farcs1$, and divided by the number of pixels and the beam size in pixels. We find that both the [\ion{C}{2}] emission and dust continuum have spatially extended emission detected at $3\sigma$ up to $\sim 8-10\ \rm{kpc}$ (note we do not correct for beam convolution given the small beam size). We fit the extended surface brightness profiles with exponentials with best-fit scale length $3.1\pm0.4\ \rm{kpc}$ and $4.1\pm0.6\ \rm{kpc}$, for the dust and [\ion{C}{2}] emission respectively, $1.5-2\times$ higher than the value found in median stacks \citep[][]{Novak2020}. In contrast to stacks however, the strength of the [\ion{C}{2}] emission in the extended profile is much higher than that of the continuum. The average continuum surface brightness at $1-2\ \rm{kpc}$ is only $\sim 0.5-1\%$ of the central peak \citep[in agreement with median stacks, e.g.][]{Novak2020}), against $\sim 5-10\%$ for the [\ion{C}{2}]. We note that the two [\ion{C}{2}] companions C1 and C2 reported in \citet{Venemans2019} at a projected sky distance of $2.0\ \rm{kpc}$ and $5.1\ \rm{kpc}$ are not detected in the continuum (neither this work nor in \citet[][]{Venemans2019}) and thus do not affect the quasar continuum emission profile (see further Appendix \ref{app:companions}). 

\begin{figure*}
    \hspace*{0.7cm}
    \includegraphics[height=0.44\textheight]{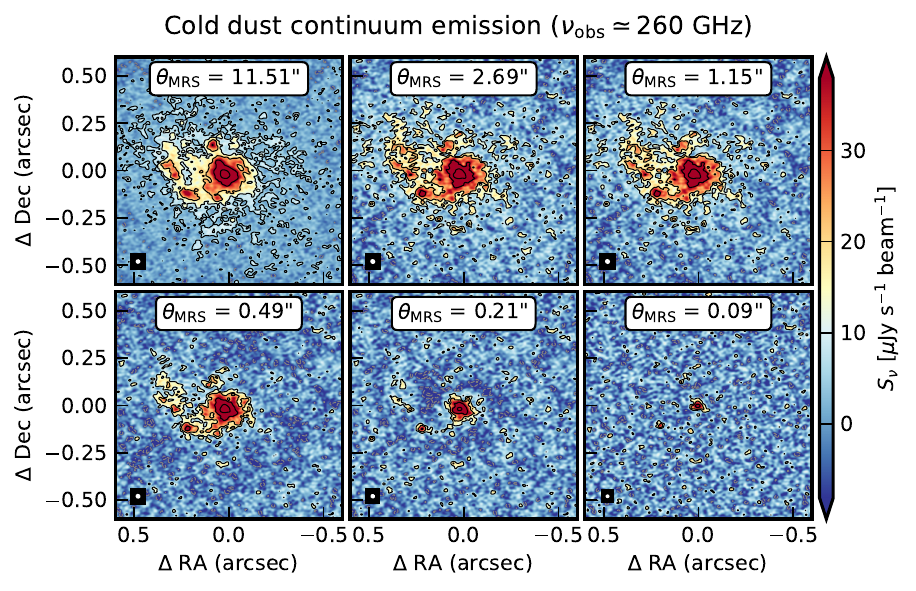} \\
    \hspace*{0.8cm}\includegraphics[height=0.44\textheight]{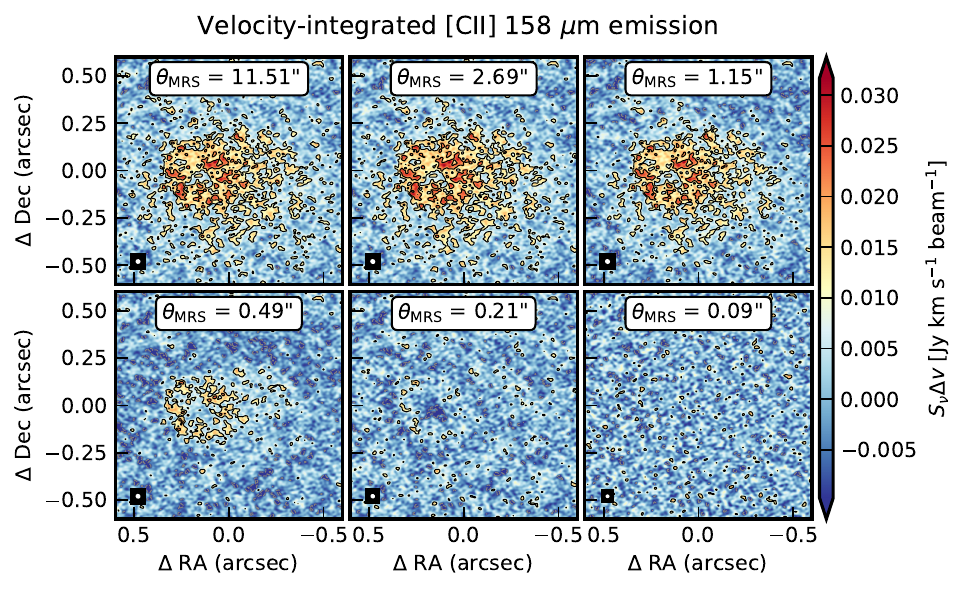}
    \caption{Cold dust continuum (top panels) and velocity-integrated [\ion{C}{2}] (bottom panels) emission of J0305--3150 as a function of the maximum-recoverable scale ($\theta_{\rm{MRS}}$). The [\ion{C}{2}] intensity declines when the MRS scales approaches $\simeq 0\farcs5$, and vanishes completely below $\sim0\farcs25$. In contrast, the dust continuum is detected with MRS scales down to $\sim 0\farcs1$, hinting at their different nature. The color scaling is identical for all [\ion{C}{2}] and dust continuum images, respectively. Contours are plotted in solid contours at the (2,4,8,16,32)$\sigma$ levels, and in dashed for negative contours at the (-2,-4)$\sigma$ levels, where $\sigma$ is the rms level in each image. The beam size is indicated with a white ellipse in the bottom left corner of each panel. }
    \label{fig:cii_dust_inneruvtaper}
\end{figure*}

Extended [\ion{C}{2}] emission (up to $\sim 10\ \rm{kpc}$) has been detected both in individual galaxies and quasars \citep[e.g.][]{Cicone2015, Ginolfi2020a_merger, Fujimoto2019a,Meyer2022a} and stacks \citep[][]{Ginolfi2020b,Bischetti2019, Novak2020}. Open questions nonetheless remain on the nature of the [\ion{C}{2}] line which could trace faint, unresolved companions, outflows, inflows \citep[][]{Fujimoto2020}, gas-stripped by mergers \citep[e.g.][]{Decarli2019,Ginolfi2020a_merger} or even shock-heated gas \citep[e.g.][]{Cormier2012, Appleton2013}. We note that similarly high resolution observations of the $z\sim 7$ quasars J2348--3054 and J0109--3047  \citep{Walter2022,Meyer2023} did not reveal extended [\ion{C}{2}] emission beyond $2-3\ \rm{kpc}$. However, the merging quasar system PJ308--21 \citep{Decarli2019} shows clear extended [\ion{C}{2}] emission, suggesting that [\ion{C}{2}] ´haloes' could be linked to gas disrupted by mergers (such as J0305--3150 and PJ308--21). Whether the gas is shock-heated or traces gas reservoirs that will fuel star-formation and SMBH growth can be determined by observations of other fine-structure lines (e.g. [\ion{O}{1}], [\ion{N}{2}]) at similarly high resolution with ALMA. 

\subsection{Characterizing the physical scale of regions emitting in the \texorpdfstring{[\ion{C}{2}]}{[CII]} line and FIR continuum}
\label{sec:physical_scales}
Having discussed the average profile and extension of the  [\ion{C}{2}] and dust emission, we now turn to the physical scales on which they are emitted. To do so, we study the [\ion{C}{2}] and dust emission directly in the UV-plane. We show the real part of the visibilities as a function of baseline length in Fig. \ref{fig:visiblities_scales} (left). The difference between the [\ion{C}{2}] and dust continuum emission is clear, with dust continuum emitted on smaller scales than [\ion{C}{2}], as also observed in the image plane (see Fig. \ref{fig:cii_dust_inneruvtaper}). Unsurprisingly, the largest scale observed (smallest baseline) in our C43-9/10 observations corresponds to the $\sim50$th percentile of the [\ion{C}{2}] distribution, in agreement with our finding that $\sim 50\%$ of the [\ion{C}{2}] flux is missing in the C43-9/10 observations compared to the more compact configuration observations (see Section \ref{sec:obs_red} and Appendix \ref{app:flux_comparisons}). 

We fit the real part of the visibilities with the cumulative density function of a log-normal distribution to recover the scales at which [\ion{C}{2}] and the dust continuum are emitted. We show the best-fit log normal distributions in Fig. \ref{fig:visiblities_scales} (right). The log-normal distribution of the dust and [\ion{C}{2}] emission have parameters $\mu_{\rm{cont}}  = 0.057\pm0.015, \sigma_{\rm{cont}}  = 0.057\pm0.15$ and $\mu_{\rm{[CII]]}}  =0.363\pm0.028,\sigma_{\rm{[CII]}}  = 0.368\pm0.026$, corresponding to physical scales of $1.5-25.4\ \rm{kpc}$ and $2.3-67.9\ \rm{kpc}$, respectively (we quote the 5-95 percentile range). The majority of the dust emission is emitted at scales consistent with that of photodissociation regions (PDRs, see Fig. \ref{fig:visiblities_scales}, right) in high-redshift SMGs $(0.5-5\ \rm{kpc})$ \citep[e.g][]{Gullberg2015, Canameras2018}. [\ion{C}{2}] is however emitted mostly at larger scales incompatible with PDR sizes. This is despite analysis of other CO, [\ion{C}{1}], and [\ion{C}{2}] line ratios in J0305--3150 showing it to be consistent with PDR excitation models \citep[][]{Venemans2017,Li2022}. Following \citet[][]{Wolfire1990, Stacey2010}, we can derive a typical scale length for the PDRs in J0305--3150 $R\propto (L_{IR}/G_0)^{1/2}$ where we use the densities ($n=10^5$) and Habing flux $10^3 < G_0 <10^4$ from \citep[][]{Venemans2017} for J0305--3150 and scale the relation to matche the measurements for M82 \citep{Joy1987}. We find $R_{\rm{PDR}}=1.2-3.8\ \rm{kpc}$, in agreement with other high-redshift SMGs and the lower tail of dust emitting scales, but not that of the [\ion{C}{2}] emission. 

It is interesting to note here that \citet[][]{Li2022} report that the CO(7-6), (6-5) and the [\ion{C}{1}] sizes match that of the dust continuum when observed at $0\farcs4$ resolution, whereas [\ion{C}{2}] is already more extended at that resolution. One can thus reasonably expect that high-resolution CO observations should trace closely the cold dense ISM of the quasar host (as seen in the dust continuum and small-scale [\ion{C}{2}]), and not the extended [\ion{C}{2}] emission tracing diffuse \ion{H}{1} gas.

For illustration purposes we demonstrate visually the different scales of [\ion{C}{2}] and dust emission in J0305--3150 by showing maps created using inner uv tapering to gradually removing large-scale emission in Fig. \ref{fig:cii_dust_inneruvtaper}. In agreement with the results of the uv-plane analysis, dust is clearly detected at $6\sigma$ around the quasar even when all baselines sensitives to scales $>$0\farcs10 have been removed, indicative of star-forming clumps on $r\lesssim 200\ \rm{pc}$ scales (see further Section \ref{sec:dust_SFRD}). In contrast, significant [\ion{C}{2}] emission starts to disappear when scales $>0.5$\arcsec\ are removed, and disappear completely when the maximum-recoverable scale $\theta_{\rm{MRS}}\simeq \ $0\farcs2 threshold is crossed. Although the analysis in the uv-plane already indicate that there is no flux emitted at small physical scales, we show further in Appendix \ref{app:flux_comparisons} how a collection of point sources (for example Giant Molecular Clouds, GMC, at the resolution of our observations) would provide a completely different [\ion{C}{2}] morphology as the one observed in J0305-3150. The fact that [\ion{C}{2}] is over-resolved in our highest-resolution observations is evidence that it is emitted in much more diffuse gas than dense PDRs around GMCs.

Given all the above, we conclude that a large fraction of [\ion{C}{2}] emission does not originate in PDR surrounding giant molecular clouds, but rather in diffuse H$_2$ or \ion{H}{1} gas. This [\ion{C}{2}]-emitted gas is located both in the interstellar medium of the galaxy but also ($\sim 50\%$) on larger scales (up to $\sim 10\ \rm{kpc}$), where the diffuse gas could be tidally stripped due to an interaction with merging companion galaxies (see Section \ref{sec:discussion_merger}).

Another important result is that we do not detect significant emission from scales $\theta \lesssim 0\farcs1$, as shown in Figure \ref{fig:visiblities_scales}, left (and inset panel). We can thus derive an upper limit on the FIR emission due to the quasar, as its emission should be point-source like and thus present in all extended baselines regardless of the MRS. Using the smallest-scales continuum map ($\theta_{MRS}=0\farcs1$), the continuum is detected in the central clump at $S_\nu = 61 \pm 6 \ \mu \rm{Jy}$, which is only $S_{\nu,QSO} / S_{\nu,total} \lesssim 1\%$ of the total continuum flux density in the quasar host of J0305--3150. Therefore the direct contribution of the quasar to the FIR continuum is negligible, although heating of the dust by the quasar could still play role \citep[e.g.][]{DiMascia2021}. We also detect another clump in continuum the same map (labeled clump B in Section \ref{sec:dust_SFRD}) with a similar flux density $S_\nu = 36 \pm 6 \ \mu \rm{Jy}$. Given that this clump is not located at the position of the quasar, this seems to suggests that the impact of AGN heating on the dust emission is limited in J0305-3150.

\section{The nature of extended \texorpdfstring{[\ion{C}{2}] 158\ $\mu\rm{m}$}{[CII] 158 um} emission}
\label{sec:kinematics}
\subsection{Kinematics of the small- and large-scale \texorpdfstring{[\ion{C}{2}]}{[CII]}-emitting gas}

Our analysis in the previous Section showed that [\ion{C}{2}] $158\ \mu\rm{m}$  is emitted by structures with physical sizes ranging from $\sim 1-10$ kpc with potentially different origins (for example an extended halo and smaller clouds in a central rotating disk). We now investigate whether  [\ion{C}{2}] shows similar kinematics at all physical scales of emission. In particular, \citet[][]{Drake2022} have shown that in $z\sim 6$ quasars, including J0305--3150, the kinematics of [\ion{C}{2}] (generally tracing the ISM of the host) and the Lyman$-\alpha$ halo (tracing the extended gas reservoir, i.e., the circumgalactic medium, CGM) are misaligned. Given that [\ion{C}{2}] seems to traces physical scales from sub-kpc to $10-20$ kpc scales in J0305--3150, we now investigate the large- and small-scale kinematics of the [\ion{C}{2}] in details. 

\begin{figure}
    \centering
    \includegraphics[width=0.48\textwidth]{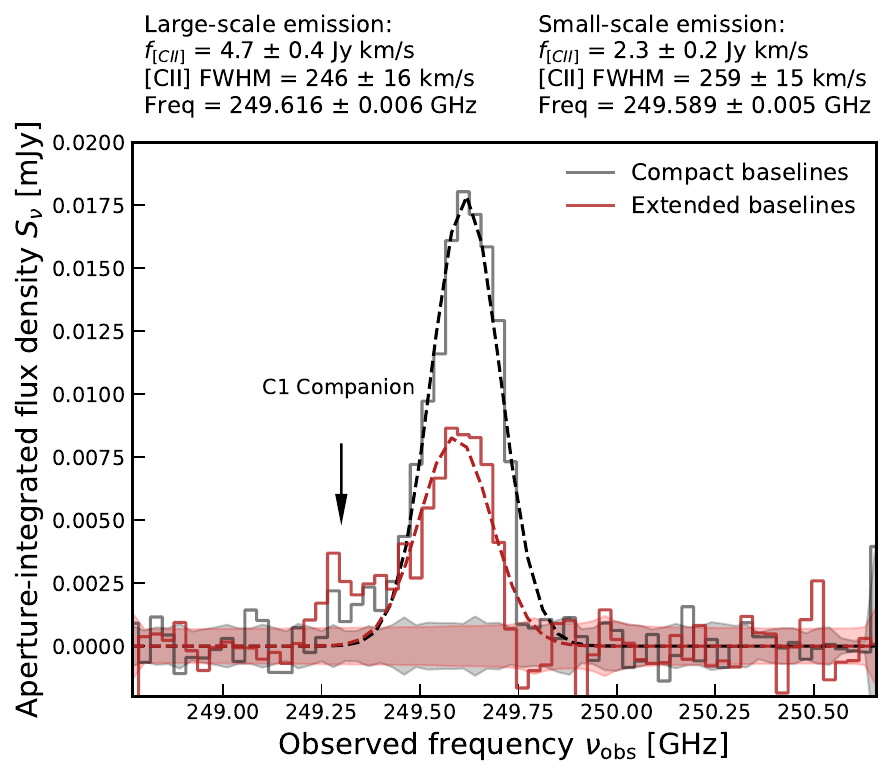}
    \caption{Aperture-integrated [\ion{C}{2}] spectra of the large-scales ($\theta>1.2$\arcsec, black) and small-scales ($\theta<1.2$\arcsec, black) [\ion{C}{2}] emission. The aperture size is $r=1\farcs5 / 0\farcs75$ for the large/small-scales emission respectively, with primary-beam and residual-scaling corrections applied. Each [\ion{C}{2}] spectrum is fitted with a simple Gaussian (dashed lines) whose mean frequency and FWHM is indicated in the upper left corner. We find no significant difference between the two [\ion{C}{2}] spectra. }
    \label{fig:spec_CII_smalllargescales}
\end{figure}

We first divide our observational data by separating baselines corresponding to observed scales of $\theta>1\farcs2$ and $\theta<1\farcs2$. We choose $\theta=1\farcs2$ to be close to the 50th percentile of the cold dust emission. The [\ion{C}{2}] is therefore not evenly split in terms of absolute fluxes, but the larger number of long baselines (see Appendix \ref{app:baselines}) results in the SNR of the two subset being roughly equal. We image the long-baselines (small-scales) data using the same procedure as described in Section (\ref{sec:obs_red}). For the large-scale emission data, we use a pixel size of $0\farcs12$ and optimal weighting (Briggs $r=0.5$) instead of natural to improve the resolution. We first present the extracted [\ion{C}{2}] spectrum of the small- and large-scale emission in Fig.\ \ref{fig:spec_CII_smalllargescales}. Both spectra have consistent central frequencies and FWHMs, and do not show any significant difference or evidence for a broad, outflowing component.

\begin{figure*}
    \centering
\includegraphics[width=1\textwidth]{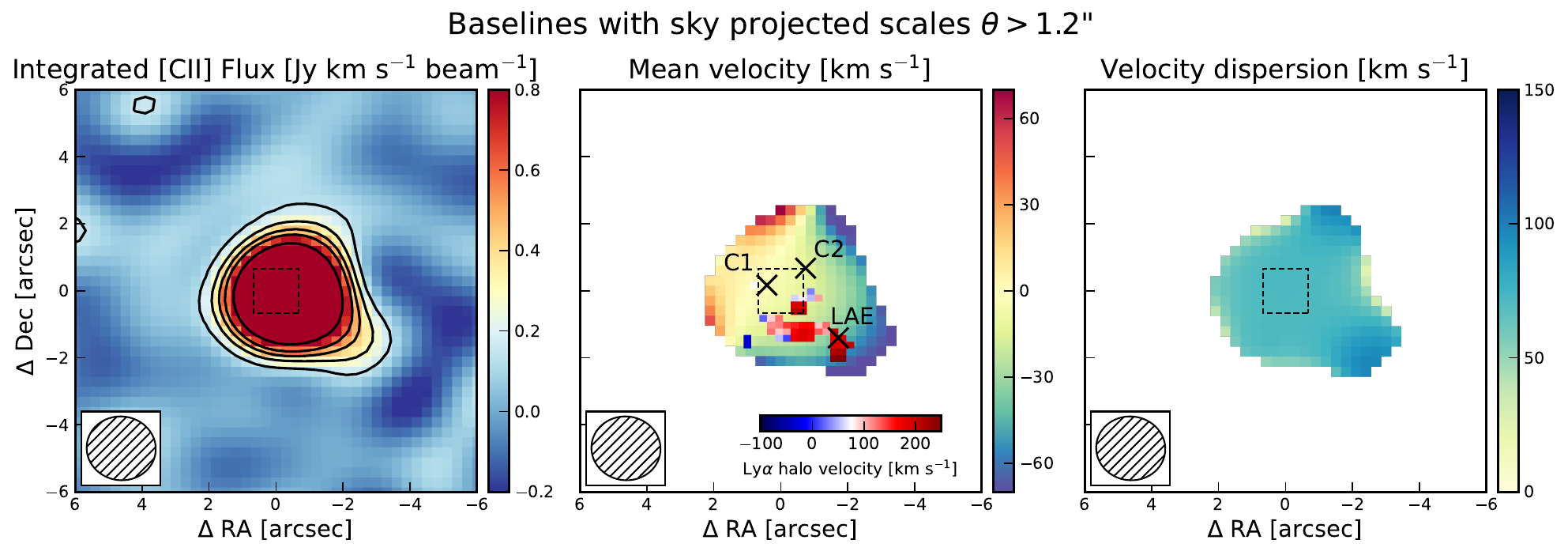}
    \includegraphics[width=\textwidth]{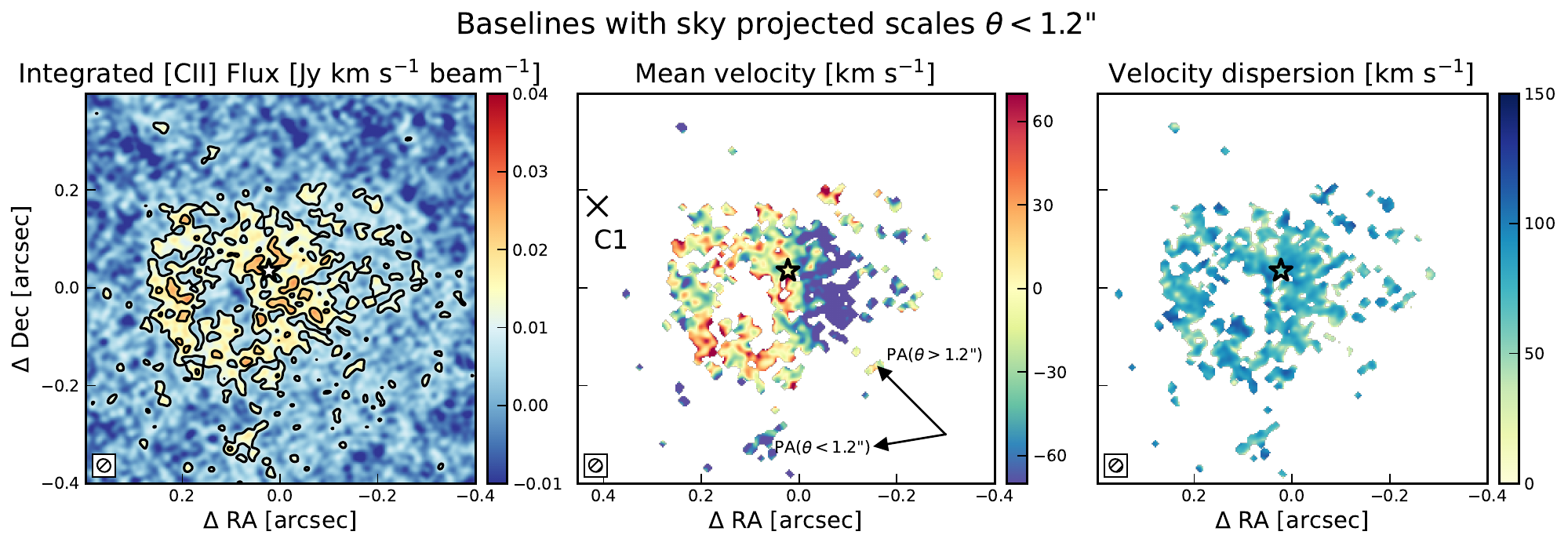}
    \caption{ [\ion{C}{2}] intensity maps and kinematics for the [\ion{C}{2}]-emitting gas at scales $>1\farcs2$ (top row) and $<1\farcs2$ (bottom row). The dashed square in the upper panels indicates the bottom row panel Field of View, illustrating the difference of scales. The rest-frame optical position of the quasar is shown with star in the bottom panels. In the top middle panel, we overlay the Lyman-$\alpha$ halo velocity map from \citet[][]{Drake2022}. The large- and small-scale [\ion{C}{2}] emission is obtained by removing the appropriate baselines directly in the uv-plane (see further text). The moment 0 maps (first column) are obtained by using all channels within $1.2\times$FWHM([\ion{C}{2}]) centered on the quasar [\ion{C}{2}] line. The moment 1 and 2 maps (second and third column) are obtained by fitting a Gaussian to each pixel in the data cube. The mask for the moment 1 and 2 maps is given by the $2\sigma$ contours of the intensity maps. }
    \label{fig:kinematics_2_scales}
\end{figure*}

We then present the velocity-integrated flux density and the kinematics maps of the small- and large-scale ($<1\farcs2$ or $>1\farcs2$, respectively) [\ion{C}{2}] emission in Fig.\ \ref{fig:kinematics_2_scales}. The two are markedly different. The small-scale [\ion{C}{2}] emission is indeed concentrated in the core of the quasar host galaxy, and matches the morphology of the dust continuum with a central emission region and a ring-like feature extending towards the North-East. The kinematics of the small-scale [\ion{C}{2}] shows some evidence for rotationally ordered motion, in agreement with previous results \citep[][]{Venemans2019, Neeleman2021}, although the [\ion{C}{2}]  distribution is complex and we could not fit it with a simple thin disk model. In contrast, the extended [\ion{C}{2}] shows marginal indication for a velocity gradient without a central increase in velocity dispersion. The position angle of the velocity gradient of the extended [\ion{C}{2}] emission differs by $\sim 45$ degrees from that of the rotating disk observed in the small-scale emission. Moreover, the large-scale velocity gradient is aligned with the axis of the ‘‘C1" companion and the ‘‘LAE" companion \citep[][and Fig. \ref{fig:kinematics_2_scales}]{Venemans2019}, as well the extended Lyman-$\alpha$ emission connecting the LAE companion and the quasar \citep[][]{Farina2017,Drake2019}. We note, however, that the velocity of the C1 and LAE companions with respect to the quasar host are $+361, +595\ \kms{}$ \citep{Venemans2019}, which is significantly higher than the velocities observed at the end of the tidal tail ($\pm 100\ \kms{}$). The apparent discrepancies with the Lyman-$\alpha$ halo and LAE velocities might be explained by the resonant nature of the Lyman$\alpha$ line and its complex radiative transfer cascade.

\subsection{A tidally-disrupted \texorpdfstring{[\ion{C}{2}]}{[CII]} companion?}
\begin{figure*}
    \centering
    \includegraphics[width=0.75\textwidth]{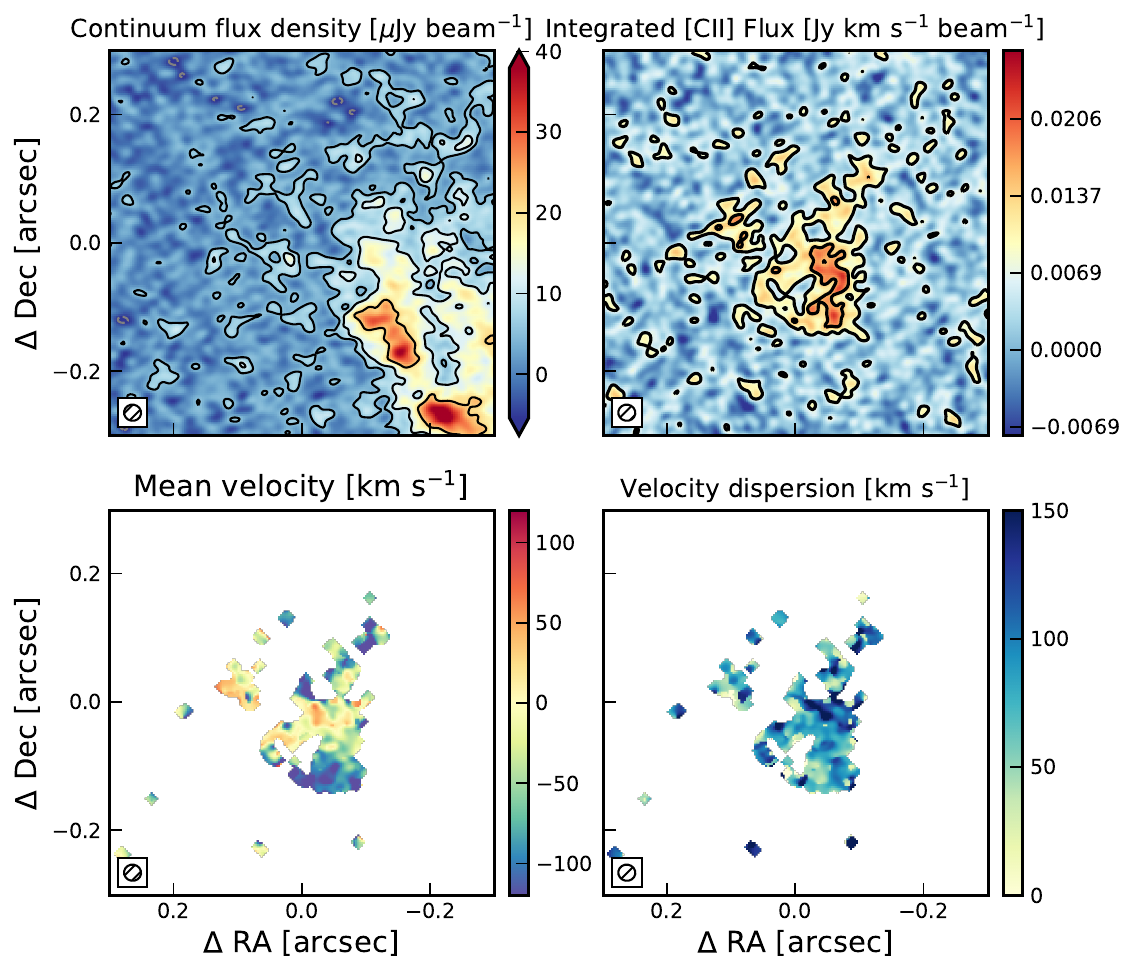}
    \caption{Top: FIR continuum at $\sim 260\ \rm{GHz}$ and velocity-integrated [\ion{C}{2}] emission of a nearby companion (named C1) of J0305--3150 discovered in \citet[][]{Venemans2019}. The companion is located at $r=2.0\ \rm{kpc}$ and $\Delta z = 0.0092 \pm 0.0003 \simeq \ 418 \rm{km\ s}^{-1}$ from the quasar host. The companion is not detected in the continuum. The contours start at $\pm2\sigma$ and increase in powers of two. The synthesized beam is plotted in the bottom left corner of each plot. Bottom: Mean velocity and velocity dispersion map of the [\ion{C}{2}] emission, computed using a Gaussian fit in pixels where [\ion{C}{2}] is detected at $>2\sigma$. The mask for the moment 1 and 2 maps is given by the $2\sigma$ contours of the moment 0, with three rounds of binary erosion and binary dilation to remove small structures due to noise (see further Section \ref{sec:obs_red}).}
    \label{fig:fig1_C1}
\end{figure*}

J0305--3150 has three companions detected in [\ion{C}{2}] in the ALMA field of view, and as well as one close Lyman-$\alpha$ emitter \citep[][]{Farina2017,Venemans2019}. The position of the companions and their continuum and [\ion{C}{2}] is detailed in Appendix \ref{app:companions}. Here we focus on the closest [\ion{C}{2}]-detected companion, dubbed C1, which is only $2.1\ \rm{kpc}$ from the quasar and therefore likely interacting with J0305--3150. We show the continuum (non-detection), integrated [\ion{C}{2}] and kinematics of the C1 companion in Figure \ref{fig:fig1_C1}. The companion exhibits a disrupted morphology, consisting of two, perhaps three, resolved [\ion{C}{2}]-emitting clumps. The kinematics are highly perturbed and complex, and we could not obtain a convincing fit with a thin disk model or a merger of two dispersion-dominated spheroids. We conclude that the companion has recently had a closed interaction with the quasar host and was consequently disrupted. In the process, it was likely tidally stripped, leading to the observed extended [\ion{C}{2}] emission. Based on the CO-based H2 masses reported by \citet[][]{Li2022} for the quasar and the companion, this is a $\sim 1:10$ minor merger. We reach a similar ratio of $\sim1:12$ using our [\ion{C}{2}] observations, but [\ion{C}{2}]-to-H$_2$ conversion factor and the overlapping [\ion{C}{2}] emission from the quasar and the companion make this estimate less precise.

\subsection{Merger-driven growth in high-redshift quasars}
\label{sec:discussion_merger}

Considering the absence of a broad outflow detected at any scales in [\ion{C}{2}], the alignment of the large-scale emission with the companions and Lyman-$\alpha$ halo, and the kinematics of the C1 companion, we therefore conclude that a large fraction (at least $50\%$) of the [\ion{C}{2}] emission is due to tidally-disrupted gas by the ongoing quasar-merger(s). The ongoing merger and the tidally-stripped [\ion{C}{2}]-emitting gas in J0305--3150 is reminiscent of the $z=6.2$ quasar PJ308--21, where \citet[][]{Decarli2019} also reported an ongoing merger. In PJ308--21 the extended bridge between the quasar and the disrupted companion [\ion{C}{2}] emission is likewise identified as tracing tidally-stripped from the ongoing merger. We speculate that eventually, the tidally-stripped gas will be accreted by the quasar host galaxy. Our observations thus support claims that mergers play an important role in replenishing the gas supply of quasar hosts \citep[e.g.][]{Lupi2022}.

This important finding is further evidence for the multi-phase nature of the gas traced by [\ion{C}{2}] in high-redshift quasars and galaxies. It also solves the [\ion{C}{2}]-Ly$\alpha$ misalignement conundrum raised by \citet[][]{Drake2022}. Our observations show that such misalignment only appears (in J0305--3150) when considering the small-scale emission [\ion{C}{2}] $<1\farcs2$ kinematics which dominate the combined dataset (e.g. Fig. \ref{fig:fig1}). Instead, the large-scale [\ion{C}{2}] emission traces kpc-scale gas similar to that detected in Lyman-$\alpha$ with VLT/MUSE, connecting the ISM of the quasar host galaxy to the larger gas reservoir at kpc scales, which is thought to be necessary to sustain both the starburst and the SMBH growth in such systems. 

\section{The resolved ISM of J0305--3150}
\label{sec:dust_SFRD}

\subsection{Integrated FIR emission, SFR and dust mass}

First of all, we present updated constraints on the integrated dust SED of J0305--3150 using all available Band 3 and 6 data as well as new ACA Band 8 observations in Appendix \ref{app:dust_sed}. The continuum observations are in agreement with previous measurements. We fit a greybody dust SED with best-fit dust temperature $T_d=25.07^{+2.27}_{-1.85}\ K$, dust emissivity index $\beta=2.28^{+0.11}_{-0.13}$, and dust mass $M_d = 4.90^{+0.17}_{-0.18} \times10^{9}\, M_\odot$ (see further Appendix \ref{app:dust_sed}). The corresponding far-infrared luminosity is $L_{FIR}= 4.03^{+0.68}_{-0.55}\times 10^{12}\ L_\odot$, and total SFR is $\rm{SFR}=507^{+55}_{-40}\ M_\odot \rm{yr}^{-1}$, assuming the \citet{Kennicutt2012} relation.

Assuming that the dust properties are constant across the host of J0305-3150, the dust mass in the brightest continuum pixel (co-spatial with the quasar rest-frame optical continuum) is $M_{d, \rm{central}} \simeq 1.1\times10^{8}\ M_\odot$. Assuming a gas-to-dust ratio of $70-100$ this largely surpasses the SMBH black hole mass ($M_{gas} \simeq 7.8-11.2\times 10^{9} \ M_{\odot}$ against $M_{BH}= 0.62\times 10^9\ M_{\odot}$, \citet[][]{Yang2023}). Thus even in the central resolution element the mass budget and kinematics are not dominated by the black hole, which is consistent with the absence of a sharp increase in the velocity dispersion at the center of J0305-3150 (see Fig. \ref{fig:fig1}).

\subsection{Star-forming clumps in J0305--3150}

\begin{table*}
    \caption{ISM properties of star-forming clumps in J0305--3150. For each region the properties are extracted with an aperture of $r=0\farcs037$ ($r= 200$ pc) centered on the regions shown in Figure \ref{fig:sfrd_map}. The FIR luminosity is derived assuming the dust properties derived for the integrated continuum of J0305--3150 (see Appendix \ref{app:dust_sed}) and scaling the dust mass for each pixel. The [\ion{C}{2}] spectra for each aperture are presented in Appendix \ref{app:cii_spectra}. }
    \centering
    \footnotesize
    \begin{tabular}{c|ccccccccc}
         & $F_{\rm{[CII]}}$ & FWHM &  $S_{\rm{cont}}$  & EW$_{\rm{[CII]}}$ & $L_{\rm{[CII]}}$  & $L_{\rm{TIR}}$  & $L_{\rm{[CII]}}/L_{\rm{FIR}}$  & $M_{dust}$& $\Sigma_{\rm{SFR (TIR)}}$ \\ 
        & [Jy km s$^{-1}$]&  [km s$^{-1}$]&   [mJy] & [$\mu\rm{m}$] &   [$10^{9}\ L_\odot$]  &  [$10^{12}\ L_\odot$]  &  [$10^{-3}$] & $[10^8\ M_\odot]$ & [$M_\odot \rm{yr}^{-1} \rm{kpc}^{-2}$]  \\   
         \hline
QSO & $0.11\pm0.01$ & $291\pm12$ & $0.404\pm0.002$ & $0.14\pm0.01$ & $0.11\pm0.01$ & $0.285\pm0.037$ & $0.404\pm0.057$ & $3.5\pm0.5$ & $286\pm37$ \\ 
A & $0.09\pm0.01$ & $201\pm9$ & $0.124\pm0.001$ & $0.37\pm0.02$ & $0.09\pm0.01$ & $0.087\pm0.011$ & $1.087\pm0.155$ & $1.1\pm0.1$ & $88\pm11$ \\ 
B & $0.08\pm0.00$ & $189\pm8$ & $0.138\pm0.002$ & $0.31\pm0.02$ & $0.09\pm0.00$ & $0.097\pm0.013$ & $0.920\pm0.130$ & $1.2\pm0.2$ & $98\pm13$ \\ 
C & $0.09\pm0.01$ & $233\pm10$ & $0.112\pm0.002$ & $0.44\pm0.03$ & $0.10\pm0.01$ & $0.079\pm0.010$ & $1.278\pm0.182$ & $1.0\pm0.1$ & $79\pm10$ \\ 
D & $0.07\pm0.01$ & $207\pm12$ & $0.101\pm0.002$ & $0.38\pm0.03$ & $0.08\pm0.01$ & $0.071\pm0.009$ & $1.115\pm0.167$ & $0.9\pm0.1$ & $72\pm9$ \\ 
E & $0.06\pm0.00$ & $171\pm7$ & $0.061\pm0.001$ & $0.55\pm0.03$ & $0.07\pm0.00$ & $0.043\pm0.006$ & $1.601\pm0.227$ & $0.5\pm0.1$ & $43\pm6$ \\ 
    \end{tabular}
    \label{tab:clumps_properties}
\end{table*}

We now study the resolved ISM of J0305--3150 at the unprecedented resolution of $140\ \rm{pc}$. In order to investigate the resolved star-formation rate density ($\Sigma_{\rm{SFR}}$), we assume that the local SFR is proportional to the continuum flux density $S_\nu$, such that at a given pixel $i$, $\rm{SFR}^i = \rm{SFR}_{\rm{tot}}\frac{ S_\nu^{i}}{\sum_j S_\nu^{j} }$ and that the $\Sigma_{\rm{SFR}}$ follows the dust continuum at $260 \rm{GHz}$\ (observed frame). This implies a constant dust temperature across J0305-3150.

\begin{figure}
    \centering
    \includegraphics[width=0.5\textwidth]{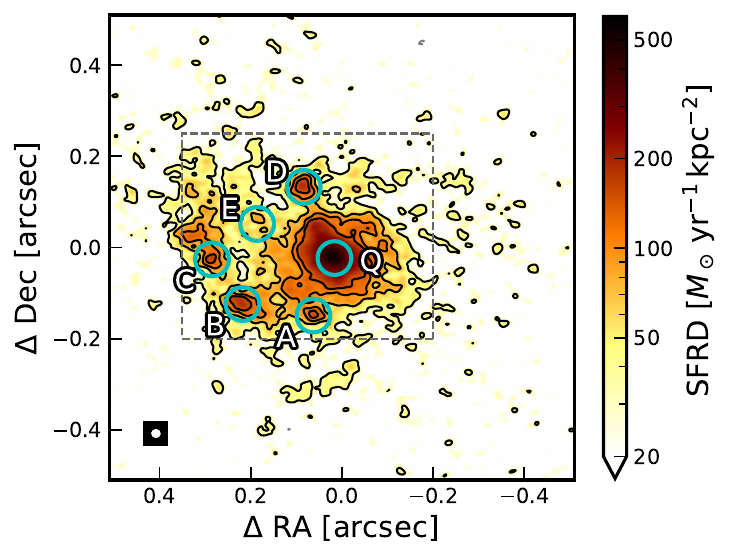}
    \caption{Star-formation rate density ($\Sigma_{\rm{SFR}}$) map of J0305--3150 assuming a proportional relation between the dust emission and the $\Sigma_{\rm{SFR}}$. The contours start at $\pm2\sigma$ and increase in steps of $2\sigma$ (up to $8\sigma$ only). The synthesized beam is plotted in the bottom left corner. The gray box shows the area tessellated with independent $r=0\farcs037\ (200\ \rm{pc}$  apertures used to study the resolved FIR properties (see further text and Fig.\ \ref{fig:resolved_CII_deficit}. The cyan circles show $r=0\farcs037$ apertures corresponding to regions of interest (see further text and Table \ref{tab:clumps_properties}). }
    \label{fig:sfrd_map}
\end{figure}

We show the resolved $\Sigma_{\rm{SFR}}$ map of J0305--3150 in Figure \ref{fig:sfrd_map}. By construction, the SFRD follows the dust emission morphology (Fig.\ \ref{fig:fig1}). The star-formation density $\Sigma_{\rm{SFR}}$ reaches up to $\sim 500\ M_\odot\ \rm{yr}^{-1} \ \rm{kpc}^{-2}$ in the central $100-200\ \rm{pc}$ surrounding the accreting SMBH. Despite being high, the maximal $\Sigma_{\rm{SFR}}$ is well below that of ´maximal'- or ´Eddington-limited'-starbursts \citep[$\rm{SFRD}_{max} \gtrsim 10^{3}-10^{4}\,M_\odot\ \rm{yr}^{-1} \ \rm{kpc}^{-2} $, e.g.][]{Thompson2005,Riechers2009,Jones2019,YangC2020,Walter2022}. We visually identify five bright continuum clumps with star-formation rate densities $\Sigma_{\rm{SFR}}$ of $\simeq 70-300\ M_\odot\ \rm{yr}^{-1} \ \rm{kpc}^{-2}$ (see Fig. \ref{fig:sfrd_map}). Their sizes are somewhat uncertain as it is difficult to disentangle the nearly beam-like emission from the underlying smooth flux profile. The effective radius of the area encompassed by the beam ($r_{eff}=(b_{\rm{min}}b_{\rm{maj}}/ (4\ln(2)))^{-1/2} = 84\ \rm{pc}$) is a reasonable approximation, but we use a slightly larger aperture to study their properties (see Table \ref{tab:clumps_properties}). For each of these clumps, we extract the continuum and [\ion{C}{2}] flux densities in apertures of $r=0\farcs037\ (r= 200\ \rm{pc}, \sim 2.4$ beams), and compute the FIR luminosity and SFR from the total SED as discussed above. Together, the four A,B,C, and D clumps accounts for $\sim8\%$ of the total FIR luminosity and, thus, SFR. When including the clump consistent with the near-infrared location of the quasar (Q), we reach $\sim15\%$ of the total FIR luminosity and SFR.  The continuum clump associated with the quasar is $\sim 3\times$ brighter than the star-forming clumps A,B,C \& D. Importantly however, we do not find significant (e.g. $>2\sigma$) difference between the star-forming clump Q and the other star-forming clumps (A,B,C,D) or the less star-forming region E in the [\ion{C}{2}]/FIR ratio, and we do not find any sign for broad [\ion{C}{2}] components indicating outflows or significantly increased velocity dispersion (see Appendix \ref{app:cii_spectra}). In summary, the continuum peak consistent with the near-infrared quasar position is simply slightly brighter in [\ion{C}{2}] and continuum than the other star-forming clumps in the host, consistent with the continuum emission being powered by star-formation.

Star-forming clumps with similar sizes and SFRDs in the rest-frame UV or optical are found in local, cosmic Noon and even higher redshift star-forming galaxies \citep[e.g.][]{Elmegreen1999,ForsterSchreiber2011,Guo2012,Smith2017,Zanella2019,Claeyssens2023, Messa2024}. Some CO observations also report detections of clumps interpreted as GMCs \citep{Swinbank2015,Tadaki2018,Dessauges-Zavadsky2019,Dessauges-Zavadsky2023}. However, FIR continuum clump detections have not been reported yet. On the contrary, studies with similar, sub-kpc observations of lower-redshift sources find a smooth FIR continuum distribution \citep[][]{Hodge2016,Gullberg2018,Rujopakarn2019,Ivison2020}, suggesting that the distribution of dust differs from that of the dense molecular gas and stars. Star-forming clumps have also been elusive in high-redshift quasars host so far, most likely due to the high resolution needed to potentially resolve them. Indeed \citet[][]{Walter2022,Meyer2023} do reach the required resolution, but only find smooth and compact [\ion{C}{2}] emission. \citet[][]{Neeleman2023} recently found one star-forming clump of similar size ($\sim 60-400\ \rm{pc}$) likely triggered by the compression of gas in a spiral arm in the host of P036+03 at $z=6.5$. In J0305--3150, the multiple star-forming clumps are more likely the result of the tidal interaction between the quasar and the companion(s), which is reminiscent of the interaction-induced clumps reported at $z\sim 2$ \citep[][]{Rujopakarn2023}. We also note that the properties of the FIR clumps detected in J0305-3150 are similar to that observed in lower-redshift galaxies in the UV/optical, despite their host galaxy hosting a luminous $\sim$ Eddington-accreting SMBH with a mass of $0.62\times 10^9\ M_{\odot}$ \citep[][]{Yang2023}. 

\subsection{The resolved \texorpdfstring{[\ion{C}{2}]}{[CII]} deficit}

\begin{figure*}
    \centering
    \includegraphics[width=0.43\textwidth]{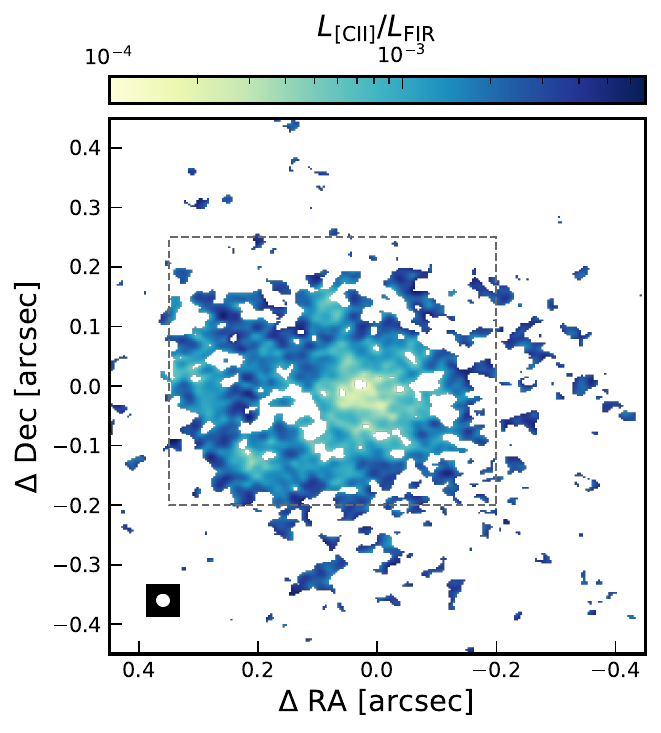}
    \includegraphics[width=0.56\textwidth]{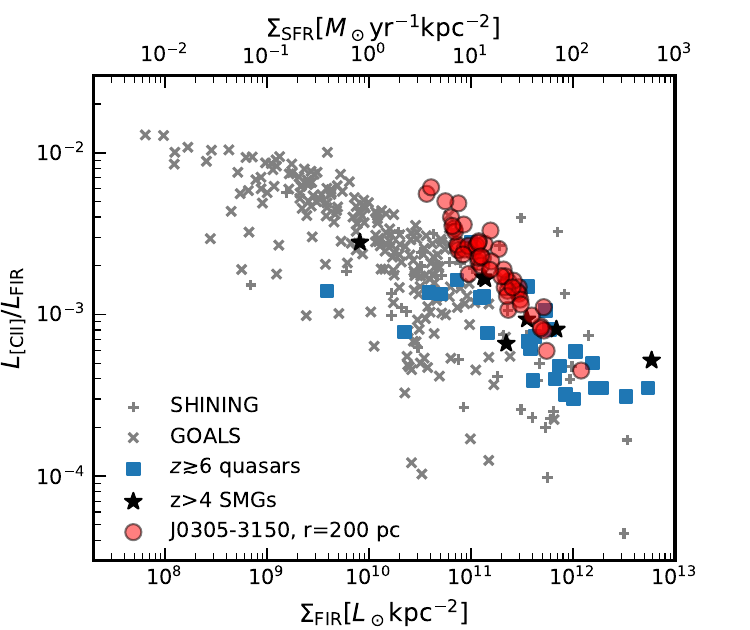}
    \caption{[\ion{C}{2}]/FIR luminosity ratio (or [\ion{C}{2}] deficit)  map (top panel) and as a function of the FIR surface density(bottom). The red circles show the values extracted in $r=200\ \rm{pc}$ in J0305--3150 (within the grey rectangle in Fig.\ \ref{fig:sfrd_map}). The grey crosses and pluses show the literature values for resolved local galaxies and ULIRGS \protect{\citep[][]{Diaz-Santos2013,Diaz-Santos2017,Herrera-Camus2018}}, the black stars for $z>4$ sub-mm galaxies \protect{\citep[][]{Walter2012,Riechers2013,DeBreuck2014,DeBreuck2019,Capak2015,Oteo2016}} and blue squares for high--redshift quasars \protect{\citep[][]{Novak2019,Venemans2020,Meyer2022a}}. The apertures in J0305--3150 follow the relation established for entire galaxies at low- and high--redshift, showing that the [\ion{C}{2}]/FIR 'deficit` relation holds down to $<200\ \rm{pc}$ scales even up to $z=6.6$. }
    \label{fig:resolved_CII_deficit}
\end{figure*}

FIR line deficits, in particular the [\ion{C}{2}] line deficit, have been important to characterise the physical conditions of local galaxies and ULIRGS. Multiple arguments have been proposed to explain the trends observed between the [\ion{C}{2}]/FIR luminosity ratio and the dust temperature of the (surface) luminosity, such as self-absorption in the [\ion{C}{2}] line, saturation of the [\ion{C}{2}] line in dense gas environments, varying dust-to-gas ratios, differential increase in the [\ion{C}{2}] and FIR luminosity with dust temperature, and/or ionisation of dust in high-UV environments (young stars or AGN) \citep[e.g.][]{Diaz-Santos2013,Diaz-Santos2017,Smith2016,Herrera-Camus2018,Herrera-Camus2018a, Lagache2018}. Importantly, both low- and high--redshift ULIRGS and SMGs follow a single trend of decreasing [\ion{C}{2}]/FIR ratio with surface FIR luminosity (see Fig.\ \ref{fig:resolved_CII_deficit}, \citet{Diaz-Santos2013,Diaz-Santos2017,Capak2015,Oteo2016}). At low-- and intermediate--redshift, \citet{Smith2017, Herrera-Camus2018, Gullberg2018} show that resolved regions (down to $\sim 200\ \rm{pc}$) also follow this trend. 

Perhaps surprisingly, high-redshift quasars host galaxies also follow the trend \citep[e.g.][]{Novak2019, Venemans2020, Meyer2022}, at least when considering unresolved and marginally resolved data. The resolution of our data enable us to test whether this is still true at smaller physical scales, especially in the few hundred parsecs around the quasar. With our $100\ \rm{pc}$--resolution data, we test whether the [\ion{C}{2}]/FIR scaling relation also applies beyond the star-forming clumps identified in J0305--3150. We take multiple independent $r=0\farcs037$ ($200\ \rm{pc}$) apertures covering the center of J0305--3150 (represented by the grey box in Fig.\ \ref{fig:sfrd_map}). We keep only those where both the continuum and the [\ion{C}{2}] emission are detected at $\rm{SNR}>2$, leaving a total of $49$ apertures. For each aperture, we compute the [\ion{C}{2}]/FIR luminosity ratio and the FIR surface brightness. We show the result in Fig.\ \ref{fig:resolved_CII_deficit} alongside the literature values of integrated galaxies. Clearly, all the regions taken in J0305--3150 follow the [\ion{C}{2}] deficit - $\Sigma_{\rm{FIR}}$ trend established at low redshift, even for those with the highest FIR luminosity close to the quasar. As a consequence, the [\ion{C}{2}] deficit relation with redshift behaves smoothly from kpc-scale unresolved observations down to $200\ \rm{pc}$ scales, mirroring the same result found for $z\sim 0$ ULIRGs, normal main-sequence galaxies and $z\sim4.5$ SMGs \citep[][]{Smith2017,Herrera-Camus2018, Gullberg2018}. We thus conclude that the [\ion{C}{2}] deficit originates from processes on $\lesssim 200 \rm{pc}$ scales, i.e. on the scales of giant molecular clouds, in a consistent fashion over 13 Gyr of cosmic history. Again, the fact that the same fine structure line deficit relations found for local main-sequence galaxies and ULIRGs also hold in luminous quasars in the first billion years is an impressive demonstration of the universality of the physical processes governing the [\ion{C}{2}]-FIR relation across cosmic time and environments.

\subsection{What is the origin of \texorpdfstring{[\ion{C}{2}]$158 \mu\rm{m}$}{[CII] 158 um}?}
We now conclude by discussing the origin of [\ion{C}{2}] $158\mu\rm{m}$ emission in J0305-3150. [\ion{C}{2}] is thought to originate mostly in PDRs. Indeed, Fine-Structure Line ratios are found to be consistent with that expected from PDR models, whether in local or distant galaxies  \citep[e.g.][]{Stacey2010, Gullberg2015,Herrera-Camus2018} or in $z\gtrsim6$ quasars \citep[e.g.][]{Novak2019,Pensabene2020,Meyer2022a,Decarli2022}. PDRs can be found in variety of regions: at the boundary layer of Giant Molecular Clouds (GMC), in the Cold/Warm Neutral Medium (CNM/WNM) and in the ionised ISM (HII regions) \citep[see][for a review]{Hollenbach1999}. In Section \ref{sec:physical_scales} (and see further Appendix \ref{app:flux_comparisons}), we have shown that [\ion{C}{2}] is emitted on scales larger than $>1\ \rm{kpc}$, excluding an origin in dense PDRs associated with GMCs. Additionally, the [\ion{N}{2}] $205\ \mu\rm{m}$/[\ion{C}{2}] ratio traces the fraction of the [\ion{C}{2}] emitted in the ionised medium \citep[e.g.][]{Oberst2011, Decarli2014, Croxall2017}. In high-redshift quasars however, the [NII]/[CII] ratio point to a subdominant contribution of the ionised phase \citep[$\lesssim 10\%$][]{Novak2019, Pensabene2021, Meyer2022a}. Having ruled out the dense molecular and ionised phase as a dominant origin for [\ion{C}{2}], we conclude that [\ion{C}{2}] $158\ \mu\rm{m}$ mostly traces diffuse atomic or molecular gas in the CNM/WNM in high-redshift quasars.

We can further hypothesize that [\ion{C}{2}] is probably not a tracer of diffuse H$_2$ gas but of atomic \ion{H}{1} in the CNM/WNM. A number of recent studies have pointed out the large discrepancies between the CO-based and [\ion{C}{2}] molecular gas masses in statistical samples of high-redshift quasars observed in multiple CO lines and [\ion{C}{2}] \citep[e.g.][]{Decarli2022,Kaasinen2024}. This is also the case in J0305-3150 where the CO and [\ion{C}{1}]-based H$_2$ gas masses \citep[($M_{H_2} = 3.0\pm1.3 \times 10^{10}\ M_\odot$)][]{Li2022} are inconsistent with the mass derived from [\ion{C}{2}] ($M_{H_2} = 1.8 \pm 0.1 \times 10^{11}\ M_\odot$) using the $\alpha_{[CII]}$ proposed by \citet[][]{Zanella2018}. The different morphologies between the star-forming clumps seen in the continuum and the [\ion{C}{2}] emission (Fig. \ref{fig:fig1}), as well as between CO and [\ion{C}{2}] \citep[][]{Shao2022,Li2022}, further supports the idea that [\ion{C}{2}] does not trace H$_2$ gas in high-redshift quasars (and potentially in other systems). Instead the joint detection of [\ion{C}{2}] and CO lines offers an interesting avenue to measure and resolve the total gas mass $M_{\rm{gas}}=M_{H2}+M_{HI}$, the atomic-to-molecular gas ratio, improved dust to gas ratios and ultimately the resolved Kennicut-Schmidt relation, provided $100-200\ \rm{pc}$-resolution observations of CO lines and the underlying continuum can be obtained.

Our conclusions are consistent with observations in low-redshift galaxies \citep[][]{Gullberg2015,Herrera-Camus2018} as well as simulations \citep[][]{Pallottini2017, Pallottini2019} that find that low-density PDRs ($n\sim 10^2-10^4\ \rm{cm}^{3}$) as found in the CNM/WNM likely produce the majority of [\ion{C}{2}] emission. Furthermore, recent simulations resolving the cold ISM conclude that [\ion{C}{2}] emission primarily comes from \ion{H}{1} gas \citep[][]{Liang2024,Casavecchia2024}. 
In this context, we interpret the detection of extended [\ion{C}{2}] haloes out to 10-15 kpc in high-redshift (quasar host) galaxies \citep[e.g.][]{Maiolino2012, Cicone2015,Carniani2017,Meyer2022a,Bischetti2019,Novak2019,Fujimoto2019a, Fujimoto2020} as [\ion{C}{2}] tracing an extended \ion{H}{1} disk or halo. In J0305-3150, the apparent connection between the large-scale [\ion{C}{2}] emission (Fig. \ref{fig:kinematics_2_scales}) and the Lyman$-\alpha$ halo supports this interpretation. Finally, \citet[][]{Schimek2024} also report that $\sim 10\%$ of the [\ion{C}{2}] emission in their simulations comes from the CGM and tidal tails associated with merging galaxies, which we hypothesized is also the case in J0305-3150. Our estimate of $\sim 50\%$ of the [\ion{C}{2}] emission originating in the diffuse extended gas following (a) merger(s) is similar to that reported in $z\sim 4-5$ galaxies \citep[][]{Ginolfi2020a_merger,DiCesare2024}.

Finally, our results suggest that multi-phase models are necessary to interpret fine-structure line and rotational lines ratios in high-redshift quasars and galaxies. Indeed, studies in the literature use a single PDR/XDR model grid for all the line ratios, implicitly relying on the hypothesis that the different lines compared (e.g. [\ion{C}{2}], [\ion{C}{1}], [\ion{O}{1}], [\ion{O}{3}], [\ion{N}{2}], CO lines) are produced in the same region \citep[e.g.][]{Venemans2017, Novak2019, Pensabene2020,Meyer2022, Li2022, Khusanova2022, Decarli2023}. The inferred hydrogen column densities $n_H \gtrsim 10^4-10^5$ are in line with that of the molecular (H$_2$) gas phase, where most of the neutral and molecular lines are likely produced ([\ion{C}{1}], [\ion{O}{1}], CO lines) but are much higher than the typical densities of the CNM/WNM that we believe [\ion{C}{2}] is mostly tracing. A re-analysis of the line ratios with multi-phase models could resolve some of the tensions reported between the various lines ratios, and in principle provide a more accurate characterization of the ISM of quasar host galaxies. 

\section{Conclusion}
\label{sec:conclusion}
We have obtained ALMA $0\farcs018$ / $97$ pc resolution observations of the [\ion{C}{2}] emission line and the underlying continuum in the $z=6.6$ quasar J0305--3150. Combined with previous observations, the nominal spatial resolution of the line and continuum imaging presented in this work is $0\farcs026 \ / \ 140\ \rm{pc}$. This is the first observation of the ISM of a galaxy in the first billion years at a spatial resolution of $\sim 100\ \rm{pc}$. At this resolution, we start to over-resolve the [\ion{C}{2}] emission, enabling us to measure the physical scales at which [\ion{C}{2}] and dust continuum are emitted. In particular, we report the following findings:

\begin{itemize}
   \item We detect little dust continuum emission emitted on scales of $\lesssim 100\ \rm{mas}$. Our observations put an upper limit on the direct contribution of the quasar to the continuum emission $S_{\nu,QSO} / S_{\nu,Host+QSO} \lesssim 1\%$ at $\nu_{rest}\simeq 2000 \ \rm{GHz}$.
   \item The [\ion{C}{2}] $158\ \mu\rm{m}$ emission and dust continuum emission do not have the same morphology. [\ion{C}{2}] is smooth and extended, whereas the dust continuum is more concentrated and clumpy. A significant fraction of [\ion{C}{2}] emission ($\sim 50\%$) is emitted in regions with sizes larger than the dust continuum, and thus likely have a different physical origin. The large-scale and small-scale [\ion{C}{2}] emission show kinematics with misaligned position angles. The small-scale [\ion{C}{2}] emission and dust continuum are emitted on scales consistent with that of PDRs and the large-scale [\ion{C}{2}] probably traces extended and diffuse \ion{H}{1} gas. 
   \item We report the detection of extended [\ion{C}{2}] emission likely due to tidal interactions with nearby companion galaxies. The large-scale, extended [\ion{C}{2}] emission is aligned with nearby companions and the offset Lyman-$\alpha$ emission detected by MUSE, with the closest [\ion{C}{2}] companion showing perturbed kinematics consistent with a post-merger state. We hypothesize that the extended [\ion{C}{2}] emission is gas tidally stripped by/from the close companion(s).
   \item We report the discovery of five star-forming clumps detected in the continuum emission with observed sizes $84\ \rm{pc} \lesssim r_{\rm{eff}}\lesssim 200\ \rm{pc}$ and SFRD $\Sigma_{\rm{SFR}} \simeq 100-300\ M_\odot\ \rm{yr}^{-1}\ \rm{kpc}^{-2}$, similar to that observed in local galaxies. The star-forming clump located at the optical position of the luminous quasar shows no different properties to the other clumps located $\sim 0.5-1\ \rm{kpc}$ away in the disk of the galaxy. Together, the five star-forming clumps account for $16\%$ of the dust-obscured SFR of J0305--3150.
   \item The $200\ \rm{pc}$ resolved [\ion{C}{2}] deficit follows smoothly the relation with the FIR surface brightness established in local and high--redshift, normal and starbursting, galaxies. The absence of a discontinuity in the vicinity of the luminous quasars indicates that the [\ion{C}{2}] deficit is not affected by the accreting SMBH but instead is driven by physical processes on $\lesssim 200\ \rm{pc}$ scales, as also found in main-sequence $z\sim 0$ galaxies \citep[][]{Smith2017}.
   \item The absence of significant [\ion{C}{2}] emission by structures with physical scale $\lesssim 1\ \rm{kpc}$ implies that [\ion{C}{2}] emission is not produced in dense PDRs located at the boundary of GMCs. We argue instead that [\ion{C}{2}] is produced in low-density PDRs in the CNM/WNM, and traces mostly diffuse \ion{H}{1} gas. This finding is in agreement with other observations and simulations of the ISM of early galaxies. It implies that [\ion{C}{2}] is a not a direct tracer of H$_2$ gas (explaining the mismatch between CO- and [\ion{C}{2}]-based gas masses in high-redshift quasars), supports the interpretation of extended [\ion{C}{2}] emission as extended (potentially CGM) gas and/or tidally-stripped gas during interactions, and motivates a multi-phase interpretation of FIR line ratios in (quasar host) galaxies.
\end{itemize}

This work shows that high resolution ($<400\ \rm{pc}$) observations of quasar host galaxies provide a unique laboratory to trace the path of gas from the extended CGM down to the accreting SMBH and the inner starbursting clumps. Further ALMA high-resolution observations of J0305--3150 can already reveal the properties of the different gas phases at $<500\ \rm{pc}$ resolution by using different gas tracers. For instance, dense molecular gas tracers such as the CO lines, should trace the dust continuum and star-formation activity, whereas ionised fine structure lines ([\ion{N}{2}] $122, 205\mu\rm{m}$, [\ion{O}{3}] $88\mu\rm{m}$) trace the ionized gas and will shed further light on the physical origin of the [\ion{C}{2}] emission. Finally, resolved dust continuum studies using the high-frequency ALMA bands will determine the dust properties and SFR on similar scales. JWST/NIRSpec IFU observations can detect the quasar host stellar light as well as the extended [\ion{C}{2}] emitting gas in the rest-frame optical line to understand its nature. Does the stellar and nebular light profile follows the cold dust continuum, the (large-scale) [\ion{C}{2}] emission, or neither of the two? 

\begin{acknowledgments}
The authors thank the anonymous referee for a detailed and thorough report which improved this work.
RAM thanks Alyssa Drake for sharing the moment 1 map of the Lyman-$\alpha$ halo of J0305-3150, Tanio Diaz-Santos for sharing the GOALS [CII]/FIR datapoints and Mirka Dessauges-Zavadsky for discussions on star-forming clumps. RAM, MN, FW acknowledge support from the ERC Advanced Grant 740246 (Cosmic\_Gas). RAM acknowledges support from the Swiss National Science Foundation (SNSF) through project grant 200020\_207349.

This paper makes use of the following ALMA data:  ADS/JAO.ALMA \#2013.1.00273.S, \#2015.1.00399.S,  
\#2017.1.01532.S,2019.1.00746.S  and 2019.2.00053.S . ALMA is a partnership of ESO (representing its member states), NSF (USA) and NINS (Japan), together with NRC (Canada), MOST and ASIAA (Taiwan), and KASI (Republic of Korea), in cooperation with the Republic of Chile. The Joint ALMA Observatory is operated by ESO, AUI/NRAO and NAOJ. 
\end{acknowledgments}
\vspace{5mm}
\facilities{ALMA}
\software{CASA \citep[][]{THECASATEAM2022}, Astropy \citep{TheAstropyCollaboration2018}, 
Numpy \citep{Numpy2020}, Matplotlib \citep{Hunter2007}, Scipy \citep{Virtanen2020}, QubeFit \citep{qubefit}, Interferopy \citep*{interferopy}}, emcee \citep[][]{Foreman-Mackey2013}

\appendix

\section{Baselines distribution and synthesised beam profile}
\label{app:baselines}
\begin{figure}[h!]
    \centering
    \includegraphics[width=0.5\textwidth]{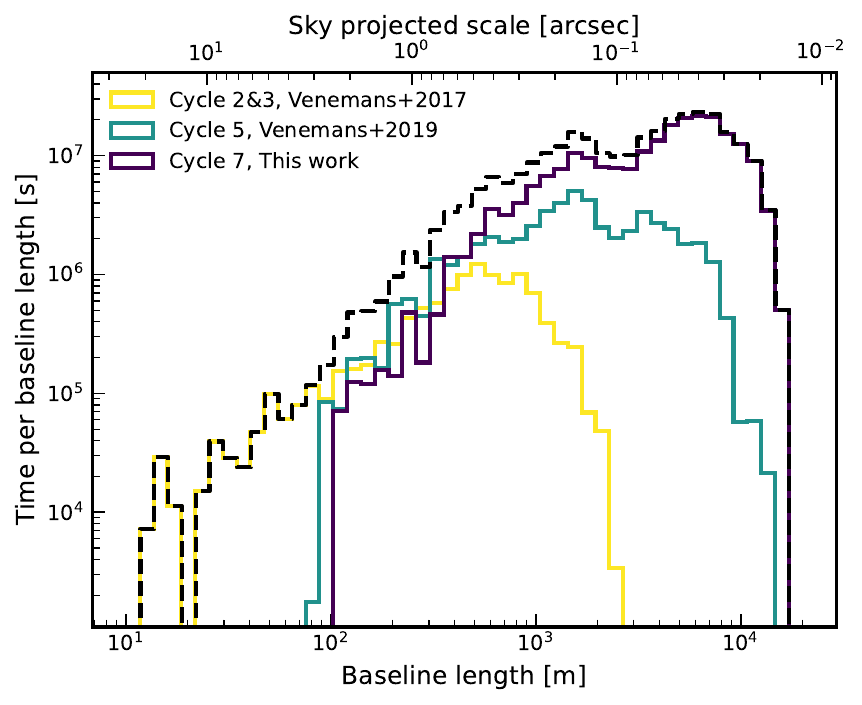}
    \caption{Distribution of time per  baseline length of the J0305--3150 ALMA observations targeting the [\ion{C}{2}] line and the underlying continuum at $\nu_{\rm{obs}}=260\ \rm{GHz}$ in J0305--3150.}
    \label{fig:baselines}
\end{figure}

We provide for reference the full distribution of baselines used in the work in Fig. \ref{fig:baselines}. This includes Cycle 2\&3 observations which first detected the [\ion{C}{2}] at $\sim0\farcs7$ resolution \citep[][]{Venemans2016}, the Cycle 5 observations at $0\farcs074, (400\ \rm{pc})$ resolution presented in \citet[][]{Venemans2019}, and the $0\farcs026$ $(140\ \rm{pc})$ observations presented in this work (see further \ref{tab:observations_summary}. The observations presented in this work not only drive the resolution down, but also significantly increase the SNR at scales $<0\farcs3$.

\begin{figure}
    \centering
    \includegraphics[height=0.31\textheight]{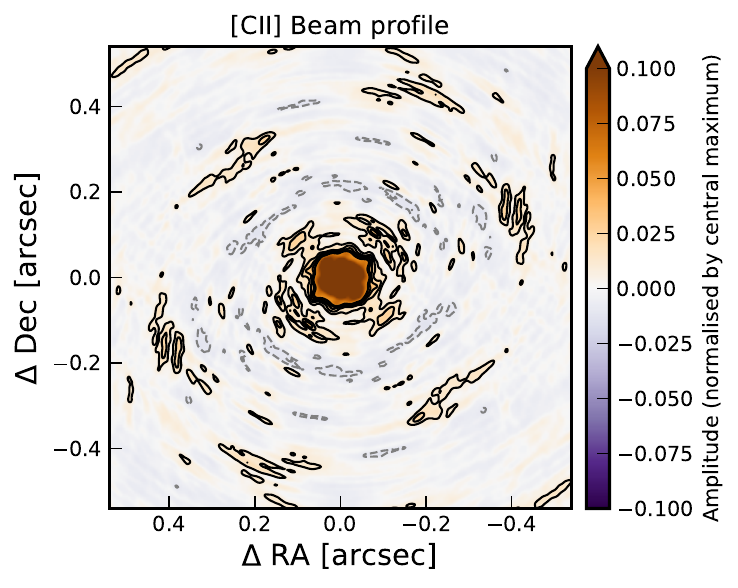} 
    \caption{Synthesized beam profile for the [\ion{C}{2}] observations. The profile is normalised to $1$. The positive (negative) contours are shown in black (gray). The contours start at $\pm 0.01$ and increase in increments of $0.01$ up to $0.05$. There is no negative sidelobe below $-0.02$, and most of the negative sidelobes are located at a radius of $\sim 0\farcs2$. Similarly, the positive sidelobes do not exceed $0.03$ and are primarily found at $r\sim 0\farcs1$ and $r\sim 0\farcs3$. }
    \label{fig:synthesizedbeam}
\end{figure}

We also show the synthesized beam profile for the [\ion{C}{2}] observations in Fig. \ref{fig:synthesizedbeam}. Notwithstanding the different array configurations combined together, the synthesized beam is well behaved. The first negative sidelobe peaks $<2\%$ of the central maximum, and the positive sidelobes at $r\sim 0\farcs1, 0\farcs3$ do not exceed $3\%$ of the beam maximum. Most importantly, the pattern and amplitude of positive or negatives sidelobes is inconsistent with any of the features seen in the cleaned images (see Fig. \ref{fig:fig1}).  

\section{Resolved-out dust continuum and \texorpdfstring{[\ion{C}{2}]}{[CII]} emission in the hyper-resolution ALMA observations}
\label{app:flux_comparisons}
In this appendix, we discuss in detail the fluxes recovered in the new $100\ \rm{pc}$ resolution observations of J0305--3150 compared to the archival $400\ \rm{pc}$ resolution data. We show that the difference between the continuum and [\ion{C}{2}] flux losses is due to the lack of sensitivity of the ALMA extended C43-9/10 configuration to the significant fraction of extended [\ion{C}{2}] emission.

\begin{figure*}
    \centering
    \includegraphics[width=0.49\textwidth]{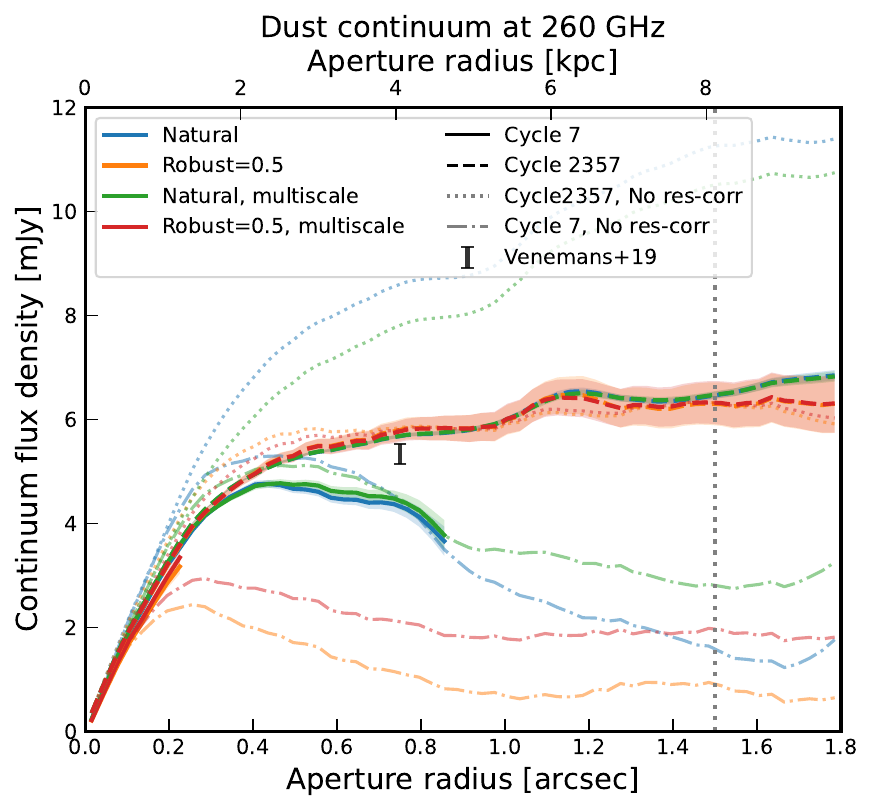}
    \includegraphics[width=0.49\textwidth]{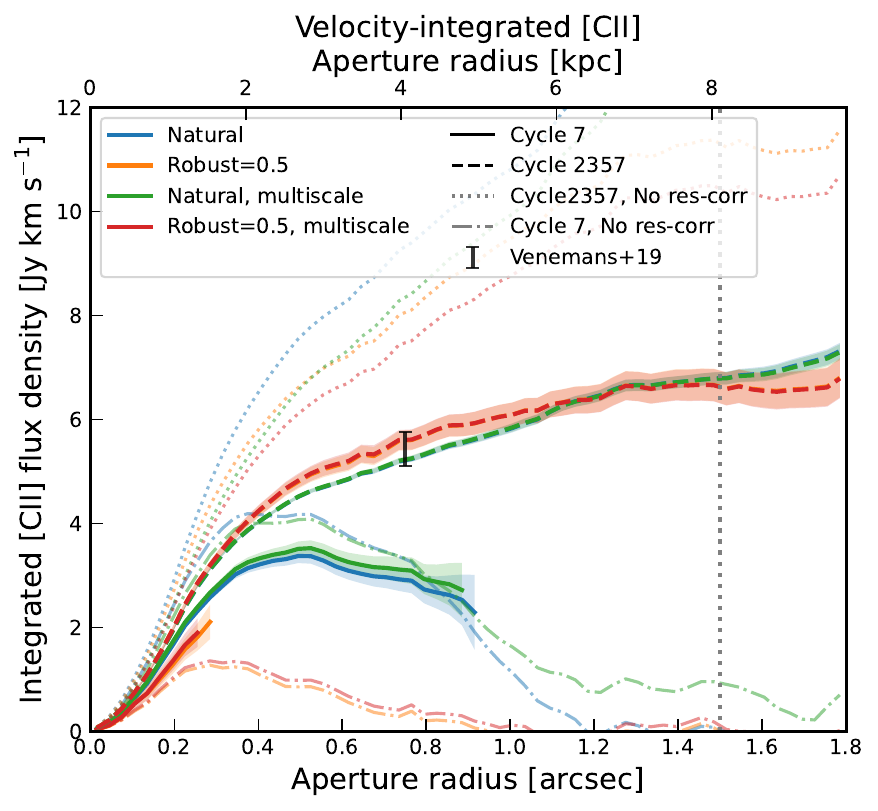}
    \caption{Aperture-integrated continuum (left) and [\ion{C}{2}] emission (right) as a function of aperture radius for different imaging schemes and datasets. The flux densities from the unresolved/partially resolved data are indicated with horizontal grey bars \citep[][]{Venemans2019}. In the highest-resolution map (Cycle 7 / C43-9/10 data only, robust=$0.5$ imaging), the absence of significant flux at $r\gtrsim 0\farcs4$ ($\gtrsim 0\farcs8$ for natural weighing) makes the residual-scaling correction unstable, hence the residual-scaled curve of growth is not plotted further. }
    \label{fig:growing_apertures}
\end{figure*}

We first compare the curve-of-growth of the continuum and [\ion{C}{2}] emission for different weightings and different dataset combinations in Figure \ref{fig:growing_apertures}. Importantly, we recover the same [\ion{C}{2}] flux as in \citet[][]{Venemans2019} using a similar aperture of $r=0\farcs75$ ($\sim 4.05\ \rm{kpc}$) . Our continuum measurement is slightly higher than reported in \citet[][]{Venemans2019} using the same aperture $r=0\farcs75$ ($5.34\pm0.19\ \rm{mJy}$ versus $5.66\pm0.03\  \rm{mJy}$). This can be explained by the fact that they report the continuum underlying the [\ion{C}{2}] at $\nu=249.5\ \rm{GHz}$, whereas we use all three continuum spectral windows to measure the continuum at an average frequency of $\nu=259.6\ \rm{GHz}$. Moreover, we report extended continuum (and [\ion{C}{2}]) up to $r\simeq 1\farcs5$ adding $10-20\%$ of flux, which was not reported in \citet[][]{Venemans2019} and \citet[][]{Li2022} (see also Fig.\ \ref{fig:cii_dust_profile}). Using a fiducial aperture of $r=1\farcs5$ and all available data (imaged with natural weighting and \textit{multiscale} cleaning) we find a total continuum flux density $S_\nu=6.45\pm0.07\ \rm{mJy}$ and a [\ion{C}{2}] flux density $6.78 \pm 0.09\ \rm{Jy\ km\ s}^{-1}$, corresponding to a line luminosity $L_{[CII]} = (7.33\pm0.10) \times 10^{9} L_\odot$. 

We find that $\sim 80-90\%$ of the low-resolution \citep[][]{Venemans2019} dust continuum emission is recovered in the Cycle 7 - C43-9/10 data, compatible with the ALMA amplitude calibration errors and the fact that they do not apply residual-scaling corrections. However, $\simeq 50\%$ of the [\ion{C}{2}] is not recovered in the C43-9/10 data with natural weighting, implying that it is partially resolved out in the C43-9/10 imaging. 

\begin{figure}
    \centering
    \includegraphics[width=0.5\textwidth]{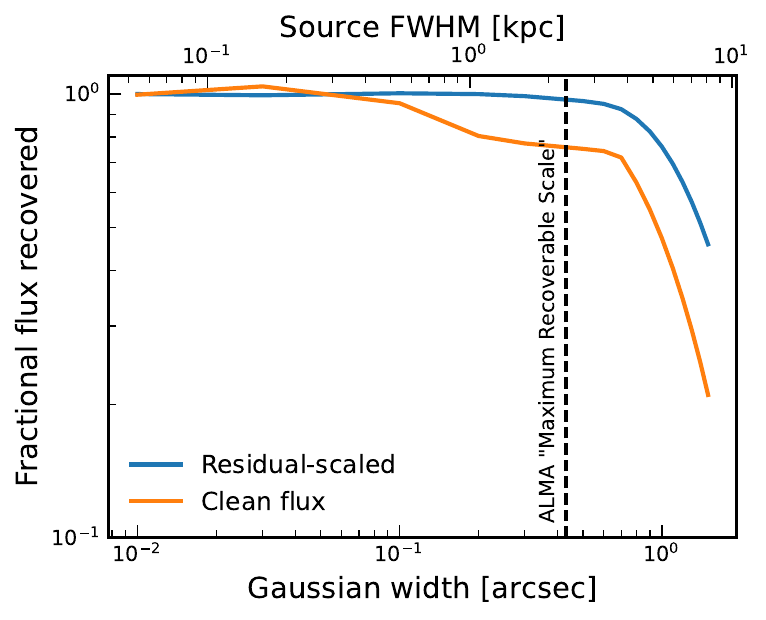}
    \caption{Fractional flux of a simulated Gaussian source with fixed spatial FWHM using the same antenna configuration ($\sim$C10) used for our C43-9/10observations of J0305--3150. The simulated visibilities are imaged with natural weighting and multiscale cleaning, as is done for the J0305--3150 data. The simulated sensitivity declines rapidly for sources with FWHM $>0\farcs7$ ($r\gtrsim3.7\ \rm{kpc}$ at z=6.6), implying that any spatial scales beyond this cannot be recovered by our C43-9/10 observations.}
    \label{fig:sensitivity_C4310}
\end{figure}

To confirm this hypothesis, we have simulated the observation of mock sources with \textit{casasim/simobserve} using the antenna configuration of our new $100\ \rm{pc}$ resolution Cycle 7 data only. The mock sources are circular 2D Gaussians with a FWHM varying between $0\farcs01-2\farcs0$ with a continuum flux density of $1\ \rm{mJy}$. The simulated visibilities (noise-free) are imaged and the source fluxes are extracted with residual scaling as following the analysis presented in this work. We show the fractional flux recovered as a function of the source FWHM in Fig.\ \ref{fig:sensitivity_C4310}. We find that the sensitivity drops precipitously for smooth sources with scale $r>0\farcs7$ ($\sim 3.8\ \rm{kpc}$ at the source redshift), which is higher than the maximum recoverable scale (MRS) reported by the ALMA archive for this project ($\theta_{MRS}=0\farcs431$). The difference stems from the definition of the MRS, which is an estimate derived from the 5th percentile of the baseline distribution.

\begin{figure}
    \centering
    \includegraphics[width=0.9\linewidth]{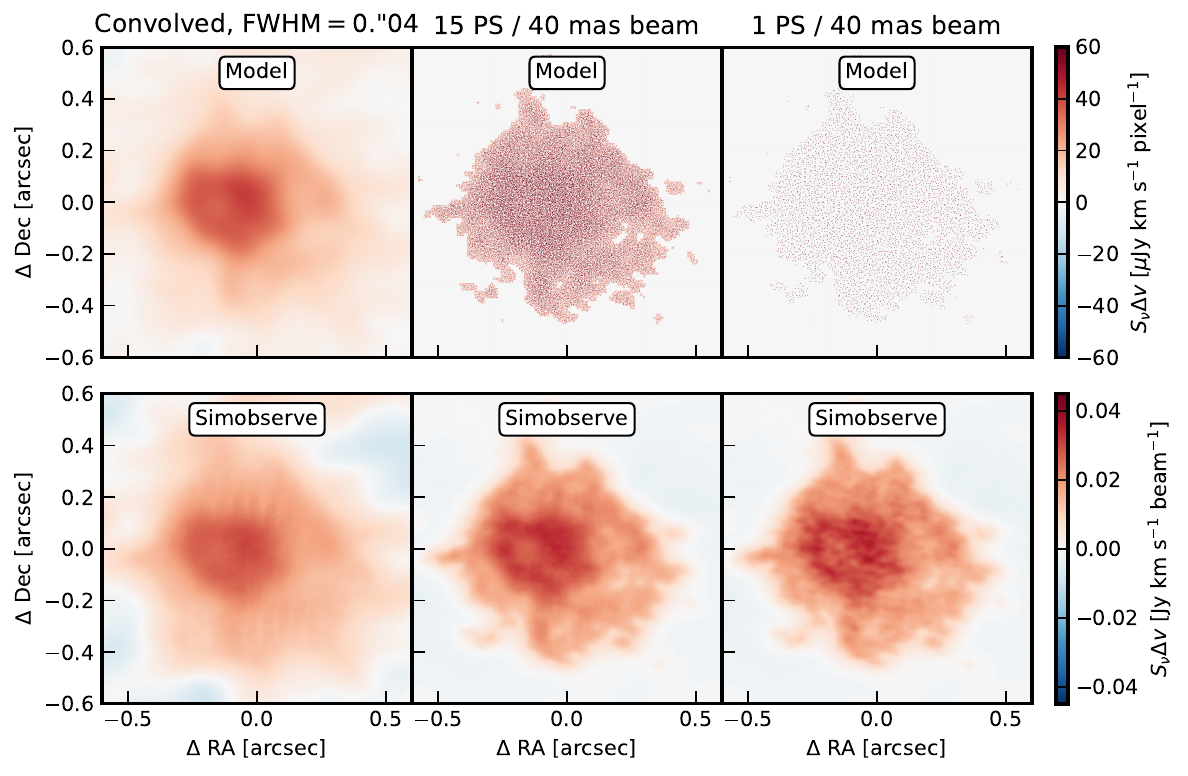}
    \caption{Mock [CII] emission morphologies (top row) and simulated observations in configuration C43-9/10 matching that of our Cycle 7 observations (bottom row). The three columns are models derived from the $\sim 75\ \rm{mas}$-resolution observations presented in \citet[][]{Venemans2019}. The first input distribution (left) is that observed b< \citet[][]{Venemans2019}, smoothed with $\sigma=40\ \rm{mas} $ (FWHM$\sim 94\ \rm{mas}$) Gaussian to remove all substructures below that scale. The second and third models (center and left), assume that the emission in each $\sim 75\ \rm{mas}$ beam can be attributed to 15 / 1 point source within the beam area. The mock observations show that a clumpy [\ion{C}{2}] distribution would not result in the smooth profile and the flux losses observed (see Fig \ref{fig:fig1}.}
    \label{fig:mock_observation_CII_morphology}
\end{figure}

We go further and show that the flux losses imply that the [\ion{C}{2}] emission cannot emerge from a collection of $<<200$ pc scale regions (such as  GMCs). We first use the observed [\ion{C}{2}] emission map at $76\ \rm{mas}$ presented in \citet[][]{Venemans2019}. We then produce three input models for CASA \texttt{simobserve}. The first model is the \citet[][]{Venemans2019} map smoothed with a FWHM$=90\ \rm{mas}$ Gaussian to remove any subtructure. The second and third models assume that the emission in each $76\ \rm{mas}$ beam is produced by 15 or 1 $\sim 10$ pc-scale source distributed randomly within the area of the beam. The three models are presented in Fig. \ref{fig:mock_observation_CII_morphology}, alongside with the mock cleaned image from \texttt{simobserve}. The color scaling is fixed for each row. We find that a more clumpy distribution of the [\ion{C}{2}] emission below the resolution of \citet[][]{Venemans2019} or our observations would result in a clumpy distribution with sharp features. Instead, our observations \ref{fig:fig1} match more closely the smoothed model, with a observed smooth [\ion{C}{2}] profile and a loss of flux relative to the other models.

\section{Revisiting the companion galaxies of J0305-3150}
\label{app:companions}
\begin{figure*}
    \centering
    \includegraphics[height=0.4\textheight]{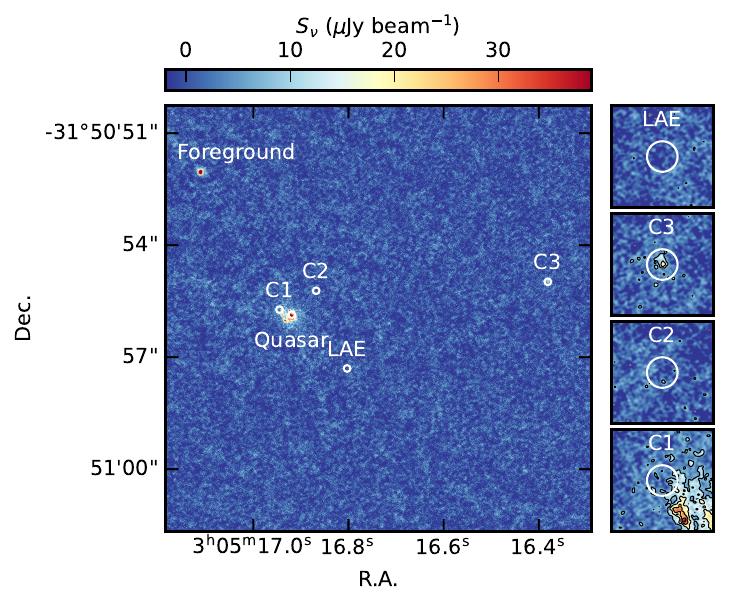}
    \includegraphics[height=0.35\textheight]{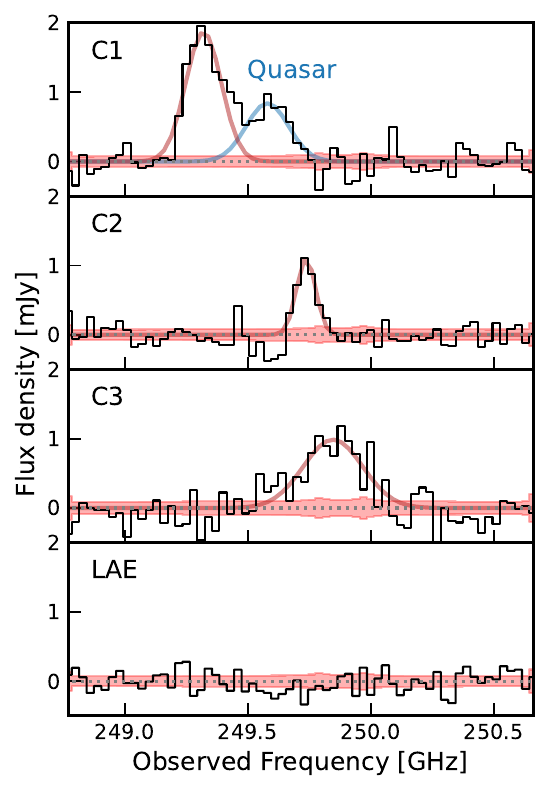} 
    \caption{\textbf{Left: } Extended map of the quasar and companions imaged in continuum at 260 GHz. \textbf{Right:} [\ion{C}{2}] spectrum of the $4$ companions. The best-fit gaussian are shown in red (we also fit the quasar emission in blue for the C1 companion) and the noise array in gray. } 
    \label{fig:companions_full}
\end{figure*}

\citet[][]{Venemans2019} reported three companion galaxies to J0305-3150 detected via their [\ion{C}{2}]. In this appendix, we detail our observations and measurements for these three objects, including one Lyman-$\alpha$ emitter reported originally in \citep[][]{Farina2017}. The continuum imaging and [\ion{C}{2}] spectrum of the companions are shown in Figure \ref{fig:companions_full}, with the measured properties in Table \ref{tab:companions}. 

We have used $r=0\farcs25$ apertures to extract both continuum and [\ion{C}{2}] spectrum at the coordinates reported in \citep[][]{Venemans2019} and apply residual-scaling corrections. For the C1 companions, close to the quasar (see Fig. \ref{fig:fig1_C1}), we fit two Gaussians to the [\ion{C}{2}] and only report the flux and FWHM for the one that is not at the quasar frequency. Only the C2 and C3 companion are detected in the aperture-integrated continuum. For C1, we give a $3\sigma$ upper limit due the lack of a clear continuum detection and the contamination of the nearby quasar. The LAE is not detected either in continuum nor in the [\ion{C}{2}]. Overall, the redshift, FWHM, and continuum fluxes are in relative agreement with \citep[][]{Venemans2019, Venemans2020}, which we note use different apertures.

\begin{table*}
    \centering
    \begin{tabular}{cccccccc}
        Name & RA & DEC & z & $r_\perp$ & $S_\nu$ & $S_{\rm{[CII]}}$ & FWHM   \\
         &  &     &   &  [kpc]  & [$\mu$Jy] & [Jy km s$^{-1}$] & [km s$^{-1}$]  \\ \hline 
       C1 &  03:05:16.95 & -31:50:55.7 & $6.62240\pm0.00022$ & $2.0^{a}$  & $<10\ (3\sigma)$ & $0.64\pm0.08$ & $228\pm24$\\ 
       C2 & 03:05:16.87 & -31:50:55.2 & $6.61011\pm0.00022$ & $5.1^{a}$  & $38\pm4$ & $0.15\pm0.04$ & $107\pm24$\\ 
C3 & 03:05:16.38 & -31:50:55.0 & $ 6.60718\pm0.00040$ & $37.4^{a}$ & $225\pm8$ & $0.65\pm0.10$ & $377\pm43$\\ 
LAE & 03:05:16.80 & -31:50:57.3 & $6.629\pm0.001^{a}$ & $11.0^{a}$ & $<10\ (3\sigma)$ & $<14.7 \ (3\sigma)$ &  -- \\ 
    \end{tabular}
    \caption{[\ion{C}{2}] and 260 GHz continuum properties of the companions in our observations. The spectrum and continuum fluxes are extracted using an aperture of $r=0\farcs25$, and residual-scaling and primary-beam correction have bene applied. For objects non-detected in [\ion{C}{2}] or continuum, we quote the $3\sigma$ rms for the [\ion{C}{2}] and continuum images (see Section \ref{sec:obs_red}).  $^{a}$ Value taken from \citet{Venemans2019}}
    \label{tab:companions}
\end{table*}

\section{Improved constraints on the total cold dust SED of J0305--3150}
\label{app:dust_sed}
In this appendix, we derive updated constraints on the total infrared luminosity and SFR of the host galaxy of J0305--3150. The dust continuum SED is constrained by ALMA observations at $100, 260$ and $400\ \rm{GHz}$ \citep[][Decarli in prep., respectively]{Venemans2017b,Venemans2019, Li2022}. However, the early low-resolution data is always incorporated in the later combined imaging and therefore the measurements are not strictly independent. We do not consider the SCUBA2 detection at $850\ \mu \rm{m}$ \citep{Li2020_sherry} given the considerably lower sensitivity and larger beam adding confusion with the numerous nearby companions and foreground galaxies reported in \citet{Venemans2019}. For each frequency, we therefore only use the latest constraints and an aperture $r=0\farcs75$: \citet[][]{Li2022} for the continuum  at $\nu_{\rm{obs}}=98.7\ \rm{GHz}$ ($S_\nu=0.27\pm0.02\ \rm{mJy}$), our improved measurement at $\nu_{\rm{obs}} = 259.6\ \rm{GHz}$ ($S_\nu=5.66\pm0.03\ \rm{mJy}$) , and finally that of Decarli et al. (in prep.) at $\nu_{\rm{obs}} = 406.9\ \rm{GHz}$ ($S_\nu=10.3\pm0.6\ \rm{mJy}$). All continuum fluxes are residual-scaled.

\begin{figure}
    \centering
    \includegraphics[width=0.53\textwidth]{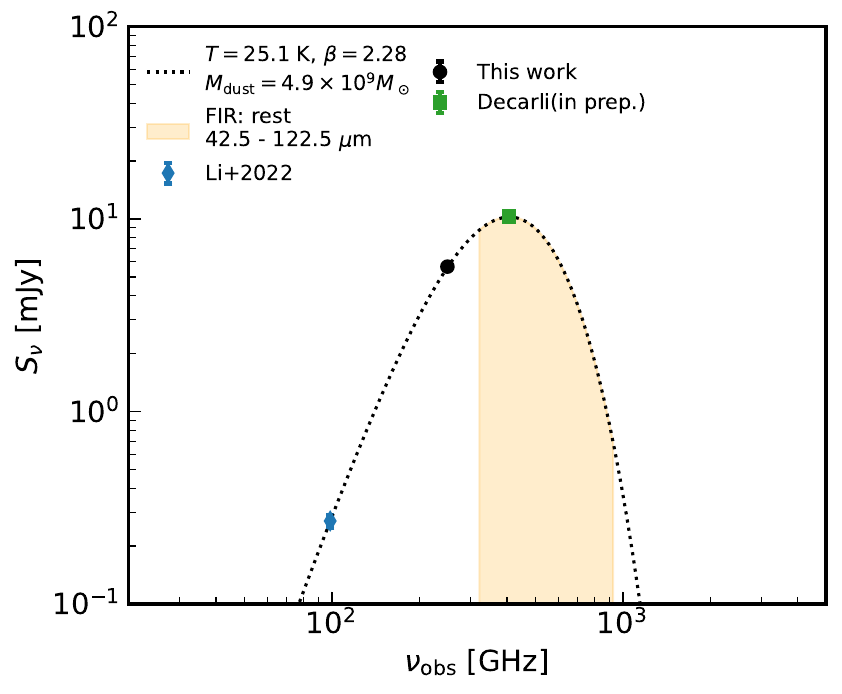} 
    \includegraphics[width=0.45\textwidth]{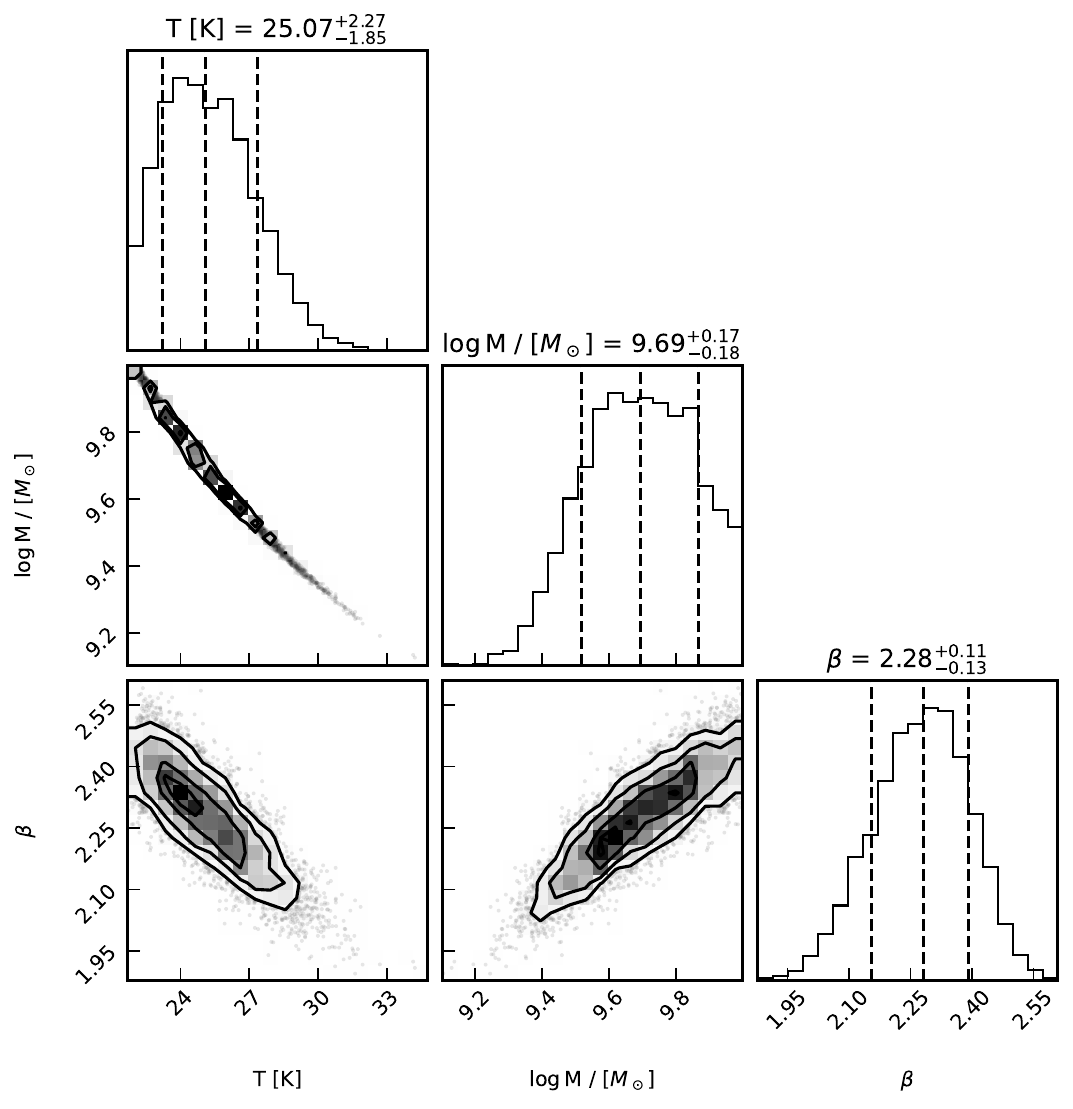}
    \caption{ \textbf{Left panel:} FIR continuum SED of J0305--3150. The dashed line shows the best-fit model. \textbf{Right panel:} Posterior distribution of the dust temperature and dust mass in the greybody dust SED model (see text for further details).}
    \label{fig:dust_sed}
\end{figure}

We model the SED using a modified blackbody spectrum with the dust emissivity index $\beta$, dust temperature $T_{d}$ and dust mass $M_d$ as free parameters. We apply the CMB contrast and heating corrections following \citet[][]{DaCunha2013}. The model parameters are well constrained with $T_d=25.07^{+2.27}_{-1.85}\ K$ and $M_d = 4.90^{+0.17}_{-0.18} \times10^{9}\, M_\odot$ and the dust emissivity index $\beta=2.28^{+0.11}_{-0.13}$ (see Fig.\ \ref{fig:dust_sed}). The total infrared luminosity ($8-1000\, \mu\rm{m}$) is $L_{TIR}= 5.79^{+0.63}_{-0.45}\times 10^{12}\, L_\odot$, the far infrared luminosity ($42.5-122.5\, \mu\rm{m}$) is $L_{FIR}= 4.03^{+0.68}_{-0.55}\times 10^{12}\ L_\odot$, and, assuming the standard \citep[][]{Kennicutt2012} relation, the total SFR is $\rm{SFR}\simeq507^{+55}_{-40}\ M_\odot \rm{yr}^{-1}$. We note that these values could be biased low by $\sim 10\%$ since we find that the continuum extends up to $1\farcs5$ (see Appendix \ref{app:flux_comparisons}), but this error is still below the statistical and systematic errors of the SFR, FIR luminosity and dust masses.

\section{\texorpdfstring{[\ion{C}{2}]}{[CII]} spectra for selected \texorpdfstring{$r=200\ \rm{pc}$}{r=200 pc} apertures}
\label{app:cii_spectra}
\begin{figure*}
    \centering
    \includegraphics[width=\textwidth,trim=0 0 6cm 0,clip]{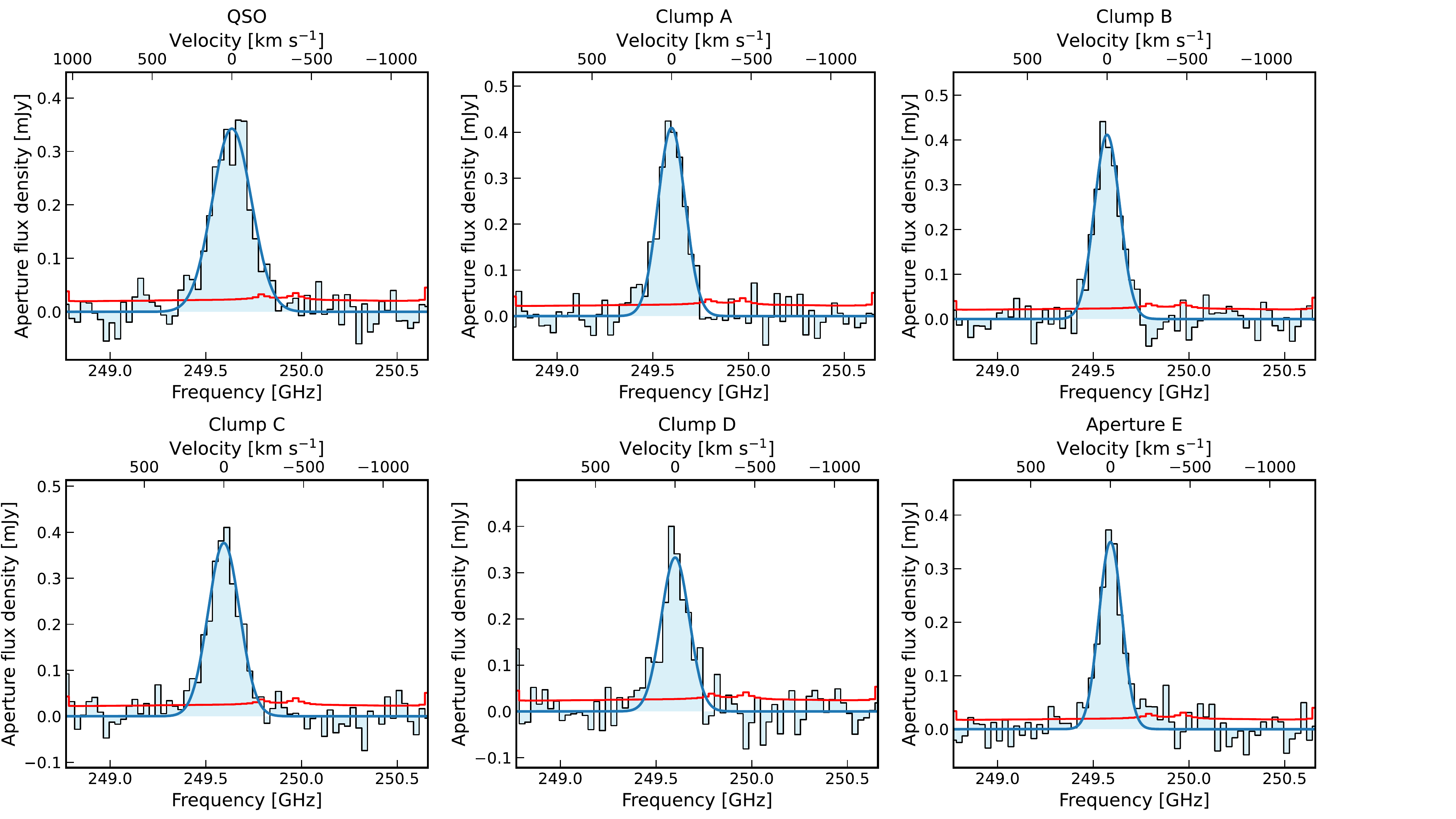}
    \caption{[\ion{C}{2}] spectra extracted from the continuum-subtracted data (all baselines) for the six $r=200\ \rm{pc}$ apertures shown in Fig.\ \ref{fig:sfrd_map}. The properties of these regions are detailed in Table \ref{tab:clumps_properties}, where residual-scaling is applied. We do not find evidence for a second broad component (potentially signalling an outflow) at the location of the quasar. }
    \label{fig:spectra_cii}
\end{figure*}

We present the [\ion{C}{2}] spectra extracted for the selected apertures shown in Fig.\ \ref{fig:sfrd_map} in Figure \ref{fig:spectra_cii}. We do not find any evidence for a high-velocity outflow at the quasar location or in any of the other selected apertures. We also provide the physical properties extracted from the 200 pc apertures tesselated over the core of J0305-3150 in Table \ref{tab:regions_properties}. 

\begin{table*}
    \caption{ISM properties of resolved $r= 200$ pc regions in the center of J0305--3150. We extract non-overlapping $r=0\farcs037$ circular aperture in the rectangular box shown in Figure \ref{fig:sfrd_map}, and only keep regions with SNR$>2$ in the dust continuum and ][\ion{C}{2}] emission. The FIR luminosity is derived assuming the dust properties derived for the integrated continuum of J0305--3150 (see Appendix \ref{app:dust_sed}) and scaling the dust mass for each pixel. }
    \centering
    \footnotesize
    \begin{tabular}{ccccccccc}
         RA & DEC & $F_{\rm{[CII]}}$ & FWHM &  $S_{\rm{cont}}$  & EW$_{\rm{[CII]}}$ & $L_{\rm{[CII]}}$  & $L_{\rm{TIR}}$  & $L_{\rm{[CII]}}/L_{\rm{FIR}}$  \\ 
        & & [Jy km s$^{-1}$]&  [km s$^{-1}$]&   [mJy] & [$\mu\rm{m}$] &   [$10^{9}\ L_\odot$]  &  [$10^{12}\ L_\odot$]  &  [$10^{-3}$]  \\   
         \hline
$46.320444$ & $-31.848917$ & $0.053\pm0.014$ & $132\pm27$ & $0.031\pm0.007$ & $0.89\pm0.31$ & $0.057\pm0.015$ & $0.022\pm0.004$ & $2.62\pm0.84$ \\ 
$46.320465$ & $-31.848917$ & $0.051\pm0.016$ & $176\pm43$ & $0.031\pm0.007$ & $0.85\pm0.33$ & $0.055\pm0.018$ & $0.022\pm0.004$ & $2.50\pm0.92$ \\ 
$46.320485$ & $-31.848917$ & $0.046\pm0.013$ & $114\pm25$ & $0.039\pm0.007$ & $0.61\pm0.20$ & $0.049\pm0.014$ & $0.028\pm0.005$ & $1.78\pm0.60$ \\ 
$46.320506$ & $-31.848917$ & $0.069\pm0.017$ & $184\pm34$ & $0.043\pm0.007$ & $0.84\pm0.24$ & $0.075\pm0.018$ & $0.030\pm0.005$ & $2.46\pm0.74$ \\ 
$46.320526$ & $-31.848917$ & $0.076\pm0.018$ & $202\pm36$ & $0.045\pm0.007$ & $0.88\pm0.24$ & $0.082\pm0.019$ & $0.032\pm0.006$ & $2.58\pm0.75$ \\ 
$46.320547$ & $-31.848917$ & $0.059\pm0.019$ & $226\pm57$ & $0.036\pm0.007$ & $0.86\pm0.33$ & $0.064\pm0.021$ & $0.025\pm0.004$ & $2.52\pm0.94$ \\ 
$46.320567$ & $-31.848917$ & $0.053\pm0.018$ & $194\pm50$ & $0.030\pm0.007$ & $0.92\pm0.37$ & $0.058\pm0.019$ & $0.021\pm0.004$ & $2.70\pm1.02$ \\ 
$46.320454$ & $-31.848899$ & $0.067\pm0.019$ & $221\pm48$ & $0.049\pm0.007$ & $0.72\pm0.23$ & $0.073\pm0.021$ & $0.035\pm0.006$ & $2.11\pm0.70$ \\ 
$46.320475$ & $-31.848899$ & $0.079\pm0.021$ & $259\pm53$ & $0.055\pm0.007$ & $0.76\pm0.22$ & $0.086\pm0.023$ & $0.038\pm0.007$ & $2.24\pm0.71$ \\ 
$46.320495$ & $-31.848899$ & $0.101\pm0.018$ & $195\pm26$ & $0.097\pm0.007$ & $0.54\pm0.10$ & $0.109\pm0.019$ & $0.068\pm0.012$ & $1.59\pm0.40$ \\ 
$46.320516$ & $-31.848899$ & $0.117\pm0.019$ & $229\pm28$ & $0.122\pm0.007$ & $0.50\pm0.09$ & $0.127\pm0.020$ & $0.086\pm0.015$ & $1.47\pm0.35$ \\ 
$46.320537$ & $-31.848899$ & $0.107\pm0.020$ & $241\pm34$ & $0.123\pm0.007$ & $0.46\pm0.09$ & $0.116\pm0.021$ & $0.087\pm0.015$ & $1.34\pm0.34$ \\ 
$46.320557$ & $-31.848899$ & $0.106\pm0.017$ & $189\pm23$ & $0.118\pm0.007$ & $0.47\pm0.08$ & $0.115\pm0.018$ & $0.083\pm0.015$ & $1.39\pm0.33$ \\ 
$46.320444$ & $-31.848882$ & $0.068\pm0.019$ & $219\pm48$ & $0.052\pm0.007$ & $0.69\pm0.21$ & $0.074\pm0.021$ & $0.037\pm0.006$ & $2.01\pm0.67$ \\ 
$46.320465$ & $-31.848882$ & $0.114\pm0.020$ & $222\pm30$ & $0.109\pm0.007$ & $0.55\pm0.10$ & $0.124\pm0.022$ & $0.077\pm0.013$ & $1.62\pm0.40$ \\ 
$46.320485$ & $-31.848882$ & $0.089\pm0.017$ & $161\pm23$ & $0.229\pm0.007$ & $0.20\pm0.04$ & $0.096\pm0.018$ & $0.161\pm0.028$ & $0.60\pm0.15$ \\ 
$46.320506$ & $-31.848882$ & $0.111\pm0.020$ & $251\pm35$ & $0.201\pm0.007$ & $0.29\pm0.05$ & $0.120\pm0.022$ & $0.142\pm0.025$ & $0.84\pm0.21$ \\ 
$46.320526$ & $-31.848882$ & $0.101\pm0.018$ & $212\pm28$ & $0.126\pm0.007$ & $0.42\pm0.08$ & $0.110\pm0.019$ & $0.089\pm0.016$ & $1.24\pm0.31$ \\ 
$46.320547$ & $-31.848882$ & $0.086\pm0.018$ & $222\pm36$ & $0.095\pm0.007$ & $0.48\pm0.11$ & $0.094\pm0.020$ & $0.067\pm0.012$ & $1.40\pm0.39$ \\ 
$46.320567$ & $-31.848882$ & $0.129\pm0.022$ & $288\pm37$ & $0.078\pm0.007$ & $0.87\pm0.16$ & $0.140\pm0.023$ & $0.055\pm0.010$ & $2.54\pm0.62$ \\ 
$46.320434$ & $-31.848864$ & $0.069\pm0.020$ & $223\pm50$ & $0.026\pm0.007$ & $1.36\pm0.52$ & $0.074\pm0.021$ & $0.019\pm0.003$ & $3.99\pm1.34$ \\ 
$46.320454$ & $-31.848864$ & $0.085\pm0.021$ & $247\pm47$ & $0.089\pm0.007$ & $0.50\pm0.13$ & $0.092\pm0.023$ & $0.063\pm0.011$ & $1.47\pm0.45$ \\ 
$46.320475$ & $-31.848864$ & $0.111\pm0.021$ & $247\pm36$ & $0.216\pm0.007$ & $0.27\pm0.05$ & $0.121\pm0.023$ & $0.152\pm0.027$ & $0.80\pm0.20$ \\ 
$46.320495$ & $-31.848864$ & $0.146\pm0.025$ & $362\pm47$ & $0.500\pm0.007$ & $0.15\pm0.03$ & $0.158\pm0.027$ & $0.352\pm0.062$ & $0.45\pm0.11$ \\ 
$46.320516$ & $-31.848864$ & $0.107\pm0.018$ & $214\pm27$ & $0.201\pm0.007$ & $0.28\pm0.05$ & $0.117\pm0.019$ & $0.142\pm0.025$ & $0.82\pm0.20$ \\ 
$46.320537$ & $-31.848864$ & $0.067\pm0.015$ & $148\pm25$ & $0.097\pm0.007$ & $0.36\pm0.08$ & $0.073\pm0.016$ & $0.068\pm0.012$ & $1.07\pm0.30$ \\ 
$46.320557$ & $-31.848864$ & $0.107\pm0.020$ & $247\pm36$ & $0.087\pm0.007$ & $0.65\pm0.13$ & $0.116\pm0.022$ & $0.061\pm0.011$ & $1.90\pm0.49$ \\ 
$46.320444$ & $-31.848846$ & $0.067\pm0.017$ & $167\pm31$ & $0.039\pm0.007$ & $0.91\pm0.27$ & $0.073\pm0.018$ & $0.028\pm0.005$ & $2.65\pm0.80$ \\ 
$46.320465$ & $-31.848846$ & $0.079\pm0.018$ & $185\pm33$ & $0.094\pm0.007$ & $0.44\pm0.11$ & $0.086\pm0.020$ & $0.066\pm0.012$ & $1.30\pm0.38$ \\ 
$46.320485$ & $-31.848846$ & $0.105\pm0.020$ & $227\pm33$ & $0.164\pm0.007$ & $0.34\pm0.06$ & $0.114\pm0.021$ & $0.116\pm0.020$ & $0.98\pm0.25$ \\ 
$46.320506$ & $-31.848846$ & $0.155\pm0.022$ & $291\pm32$ & $0.215\pm0.007$ & $0.38\pm0.05$ & $0.168\pm0.024$ & $0.151\pm0.027$ & $1.11\pm0.25$ \\ 
$46.320526$ & $-31.848846$ & $0.101\pm0.018$ & $198\pm26$ & $0.090\pm0.007$ & $0.59\pm0.11$ & $0.109\pm0.019$ & $0.064\pm0.011$ & $1.72\pm0.43$ \\ 
$46.320547$ & $-31.848846$ & $0.092\pm0.016$ & $175\pm24$ & $0.082\pm0.007$ & $0.59\pm0.11$ & $0.100\pm0.018$ & $0.058\pm0.010$ & $1.72\pm0.43$ \\ 
$46.320567$ & $-31.848846$ & $0.101\pm0.020$ & $257\pm39$ & $0.106\pm0.007$ & $0.50\pm0.10$ & $0.110\pm0.022$ & $0.075\pm0.013$ & $1.47\pm0.39$ \\ 
$46.320434$ & $-31.848828$ & $0.063\pm0.016$ & $158\pm30$ & $0.028\pm0.007$ & $1.16\pm0.40$ & $0.068\pm0.017$ & $0.020\pm0.004$ & $3.40\pm1.04$ \\ 
$46.320454$ & $-31.848828$ & $0.099\pm0.024$ & $353\pm66$ & $0.031\pm0.007$ & $1.66\pm0.54$ & $0.107\pm0.026$ & $0.022\pm0.004$ & $4.87\pm1.45$ \\ 
$46.320475$ & $-31.848828$ & $0.101\pm0.024$ & $344\pm64$ & $0.057\pm0.007$ & $0.93\pm0.25$ & $0.109\pm0.026$ & $0.040\pm0.007$ & $2.74\pm0.81$ \\ 
$46.320495$ & $-31.848828$ & $0.138\pm0.028$ & $470\pm73$ & $0.065\pm0.007$ & $1.13\pm0.26$ & $0.150\pm0.030$ & $0.045\pm0.008$ & $3.30\pm0.88$ \\ 
$46.320516$ & $-31.848828$ & $0.096\pm0.017$ & $179\pm24$ & $0.128\pm0.007$ & $0.39\pm0.07$ & $0.104\pm0.018$ & $0.090\pm0.016$ & $1.15\pm0.29$ \\ 
$46.320537$ & $-31.848828$ & $0.081\pm0.019$ & $232\pm41$ & $0.047\pm0.007$ & $0.91\pm0.25$ & $0.088\pm0.021$ & $0.033\pm0.006$ & $2.66\pm0.77$ \\ 
$46.320557$ & $-31.848828$ & $0.091\pm0.017$ & $186\pm27$ & $0.066\pm0.007$ & $0.73\pm0.16$ & $0.099\pm0.019$ & $0.046\pm0.008$ & $2.13\pm0.55$ \\ 
$46.320444$ & $-31.848810$ & $0.061\pm0.017$ & $188\pm40$ & $0.029\pm0.007$ & $1.11\pm0.40$ & $0.067\pm0.018$ & $0.020\pm0.004$ & $3.26\pm1.07$ \\ 
$46.320465$ & $-31.848810$ & $0.083\pm0.017$ & $201\pm32$ & $0.035\pm0.007$ & $1.23\pm0.35$ & $0.089\pm0.019$ & $0.025\pm0.004$ & $3.60\pm0.98$ \\ 
$46.320485$ & $-31.848810$ & $0.092\pm0.021$ & $308\pm55$ & $0.050\pm0.007$ & $0.97\pm0.26$ & $0.099\pm0.023$ & $0.035\pm0.006$ & $2.84\pm0.83$ \\ 
$46.320506$ & $-31.848810$ & $0.078\pm0.019$ & $253\pm48$ & $0.064\pm0.007$ & $0.64\pm0.17$ & $0.085\pm0.021$ & $0.045\pm0.008$ & $1.89\pm0.57$ \\ 
$46.320526$ & $-31.848810$ & $0.071\pm0.022$ & $304\pm72$ & $0.049\pm0.007$ & $0.77\pm0.26$ & $0.078\pm0.024$ & $0.035\pm0.006$ & $2.24\pm0.79$ \\ 
$46.320547$ & $-31.848810$ & $0.088\pm0.025$ & $414\pm92$ & $0.049\pm0.007$ & $0.95\pm0.30$ & $0.096\pm0.028$ & $0.035\pm0.006$ & $2.77\pm0.93$ \\ 
$46.320567$ & $-31.848810$ & $0.077\pm0.026$ & $451\pm120$ & $0.052\pm0.007$ & $0.78\pm0.29$ & $0.084\pm0.029$ & $0.037\pm0.006$ & $2.29\pm0.88$ \\ 
$46.320475$ & $-31.848793$ & $0.055\pm0.021$ & $292\pm84$ & $0.015\pm0.007$ & $1.90\pm1.10$ & $0.060\pm0.023$ & $0.011\pm0.002$ & $5.57\pm2.31$ \\ 
$46.320495$ & $-31.848793$ & $0.063\pm0.023$ & $359\pm102$ & $0.027\pm0.007$ & $1.20\pm0.53$ & $0.068\pm0.025$ & $0.019\pm0.003$ & $3.52\pm1.44$ \\ 
$46.320516$ & $-31.848793$ & $0.067\pm0.027$ & $445\pm137$ & $0.017\pm0.007$ & $2.09\pm1.17$ & $0.073\pm0.029$ & $0.012\pm0.002$ & $6.11\pm2.65$ \\ 
$46.320537$ & $-31.848793$ & $0.076\pm0.026$ & $418\pm112$ & $0.023\pm0.007$ & $1.71\pm0.77$ & $0.082\pm0.028$ & $0.016\pm0.003$ & $5.01\pm1.94$ \\ 
$46.320557$ & $-31.848793$ & $0.053\pm0.020$ & $289\pm84$ & $0.035\pm0.007$ & $0.80\pm0.34$ & $0.057\pm0.022$ & $0.024\pm0.004$ & $2.35\pm0.99$ \\ 
    \end{tabular}
    \label{tab:regions_properties}
\end{table*}

\bibliography{bib}{}
\bibliographystyle{aasjournal}

\end{document}